\documentclass[prd,twocolumn,showpacs,nofootinbib]{revtex4}
\usepackage[dvips]{graphicx}
\usepackage{amsmath}

\usepackage{feynmp}

\newcommand{\rh}{\rho}
\newcommand{\ph}{\phi}
\newcommand{\ps}{\psi}
\newcommand{\lm}{\lambda}
\newcommand{\om}{\omega}
\newcommand{\veck}{{\bf k}}

\newcommand{\bhd}{\hat{b}^{\dagger}}
\newcommand{\vac}{|0\rangle}
\newcommand{\vecx}{{\bf x}}

\newcommand{\mf}{\text{mf}}
\newcommand{\class}{\text{class}}

\newcommand{\omlat}[1]{\omega_k^{(#1)}}
\newcommand{\re}{\Re}
\newcommand{\symmetric}{``symmetric''}
\newcommand{\symm}{``symm''}
\newcommand{\broken}{``broken''}
\newcommand{\symmetricph}{``symmetric phase''}
\newcommand{\brokenph}{``broken phase''}

\begin{document}

\preprint{ITFA-02-35}

\title{The Hartree ensemble approximation revisited: The \symmetricph{}}

\author{Mischa~Sall\'e}
\email{mischa.salle@helsinki.fi}

\altaffiliation[current address: ]{Helsinki Institute of Physics, P.O. Box 64,
FIN-00014 University of Helsinki, Finland}

\author{Jan~Smit}
\email{jsmit@science.uva.nl}

\affiliation{Institute for Theoretical Physics, University of Amsterdam\\
 Valckenierstraat 65, 1018 XE Amsterdam, the Netherlands}

\date{\today}

\begin{abstract}
The Hartree ensemble approximation is studied in the \symmetricph{} of $1+1$
dimensional $\lambda \phi^4$ theory. In comparison with the \brokenph{}
studied previously, it is shown that the dynamical evolution of observables
such as the particle distribution, energy exchange and auto-correlation 
functions,
is substantially slower. Approximate thermalization is found only for
relatively large energy densities and couplings.
\end{abstract}

\pacs{11.10.Wx, 05.70.Ln, 11.15.Ha}

\maketitle
\section{Introduction \label{sec:intro}}

Non-perturbative treatments of relativistic quantum fields out of equilibrium
are currently under intense investigation, because of their importance in
applications to the physics of the early universe and to heavy-ion collisions.
The Hartree approximation is sufficiently simple that it facilitates the study
of inhomogeneous systems by numerical methods. The possibility of inhomogeneity
allows for the incorporation of non-perturbative configurations such as
sphalerons and Skyrmions, or kinks in 1+1 dimensions.

Another aspect of inhomogeneity relates to equilibration and thermalization. For
homogeneous systems, the usual Hartree approximation fails to describe
thermalization, because it does not incorporate direct scattering. On the
other hand, the classical approximation, in which expectation values are
obtained as an average over realizations which are specified by an initial
distribution, does allow for (classical) thermalization \cite{AaBo00}. The
reason is that the realizations are typically inhomogeneous, even if the
initial distribution is homogeneous, which allows for scattering of the
classical
waves. (Clearly, strictly homogeneous classical scalar-field configurations 
cannot thermalize as these correspond to a dynamical system with only one 
degree of freedom.) Classical waves that are small 
fluctuations on a ground state, correspond in the 
quantum analogue to a homogeneous mean field with a relatively small
(``vacuum size'') two-point function. The
classical scattering of such waves would correspond to direct scattering
in the quantum analogue, which is absent in the Hartree approximation.
However, one may doubt, if 
strongly non-linear classical waves can be very well represented, 
in their quantum analogue, by two-point functions with,  
necessarily approximate,
direct-scattering terms in their evolution equations. 
One can furthermore imagine that in such situations strong non-linearity 
is important for the first 
stages of equilibration, and that this aspect could be better represented 
by strong and typically inhomogeneous mean fields.

Motivated by the capability of the classical description to incorporate
strong non-linearity,
we recently introduced \cite{SaSm00a,SaSm00b} a Hartree-ensemble
approximation, in which the initial density matrix is expanded on a basis
of Gaussian coherent states,
\begin{equation}
\hat\rh = \int D\ph\, D\pi\, \rh[\ph,\pi]\, |\ph\pi\rangle\langle\ph\pi|,
\label{eq:HEA}
\end{equation}
with ``mean fields'' $\ph$ and $\pi = \dot\ph$ and suitable variances.
With ``mean field'' we here denote the expectation value in the Gaussian
quantum density matrix $|\ph,\pi\rangle\langle\ph,\pi|$. The true mean fields
are obtained after taking also the classical average over the separate
``realizations'', weighted with the classical density functional $\rh[\ph,\pi]$.
For each of these ``realizations'' separately we use the Hartree
approximation.
 
This formulation achieves two things.
First, it allows for the description of systems with
non-Gaussian initial density matrices, thereby going beyond the Hartree
approximation, and second, it allows for strongly non-linear equilibration.
There is also a remnant of the ``small fluctuation type'' scattering:
the ``mean fields'' in the ``realizations''
$|\ph\pi\rangle\langle\ph\pi|$ are in general inhomogeneous, even if the
full expectation values describe a homogeneous system,
and these inhomogeneities
lead to indirect scattering of the quantum modes via the Fourier modes of the
inhomogeneous ``mean field''.\footnote{Of course, we would like to include 
direct scattering as well, as in \cite{BeCo01}, but this is at present 
numerically too demanding in the inhomogeneous case.}
  
Note that when the initial distribution $\rho[\phi,\pi]$ is positive,
the initial correlation functions can in principle be calculated to 
arbitrary precision using Monte Carlo methods.
However, the dynamical equations of motion still apply to an individual Gaussian
``realization'', using a Gaussian approximation (Hartree), 
and the correlation functions at later times depend on these approximations. 
When the ``mean field'' dynamics in the ``realizations'' dominates over the
Hartree corrections, we expect that this Hartree-ensemble method works better
than the usual Hartree method in terms of the true mean field, especially
in case the latter is homogeneous.

In Ref.~\cite{SaSm00a,SaSm00b} we studied this
extension of the usual Hartree approximation
in the 1+1 dimensional $\phi^4$ model,
and found that it leads to approximate thermalization.
We focused on the \brokenph{} of the system in these papers. In
the present work we discuss the \symmetricph{}.\footnote{At
zero temperature the expectation value of the order field $\phi$
distinguishes the phase regions in coupling-parameter space. At finite
temperature this expectation value vanishes
due to kink-antikink condensation. However,
there is still a clear distinction in the behaviour of the system
in the two phases even in finite volume.}

The reason for presenting results also in the \symmetricph{} is twofold.
First, it is interesting in its own right. Not much is known yet
about inhomogeneous relativistic systems out of equilibrium and numerical
data can help to sharpen our intuition.
Second, results by Bettencourt et al.~\cite{BePa01}, which were obtained
in the \symmetricph{},
appear to contradict our earlier results in \cite{SaSm00a,SaSm00b}.
We try to clarify the situation here, and substantiate our earlier
observation
that the evolution towards equilibrium in the \symmetricph{} is
much slower than in the \brokenph{}.

Broadly speaking, our results are as follows.
We find (relatively) rapid approximate thermalization from an initial 
homogeneous non-equilibrium density matrix, in case of
{\em high} energy-densities. We interpret this as concording our intuition on
strong ``mean fields''. On the other hand, for {\em low} energy densities
and/or couplings the 
time scales for reasonable changes in particle distribution functions 
are huge, even larger than what we found earlier in the \brokenph{}, which
makes numerical study very difficult. 
The inhomogeneous case studied by Bettencourt et al.~\cite{BePa01} 
(a single Gaussian wave packet) falls in the low energy density class 
and indeed, even after very large times, we do not find thermalization. 
A more complete summary of the many detailed 
results will be given at the end of this paper.

In Sec.~\ref{sec:eq_ini} we will briefly review the equations of motion,
explain the Hartree ensemble approximation, describe various initial
conditions and observables, and describe some aspects of the elegant
definition of the theory on a space-time lattice.
In Sec.~\ref{sec:numres} we will present results obtained using
numerical simulations.
We give an
explanation of a phenomenon
found in the simulations that we call ``local $k$-space equilibration''
in Sec.~\ref{sec:scatter},
where we also comment on the difference between the \symmetric{} and 
\brokenph{}.  A discussion is in Sec.~\ref{sec:discus}.
Finally in
the Appendix we derive the initial as well as free-field
form of the particle distribution for the Gaussian wave packet.

In the rest of this paper we will drop the quotes
around ``mean field''.

\section{Hartree approximations, initial conditions and
observables\label{sec:eq_ini}}

\subsection{Hartree approximation \label{subsec:eom}}

For the $\lambda \phi^4$ theory in 1+1 dimensions
the Heisenberg equations of motion are given by
\begin{equation}
(- \partial^2 + \mu^2)\hat{\phi}(x) + \lambda\hat{\phi}(x)^3 = 0.
\label{eq:opeq}
\end{equation}
In the Hartree approximation (H.a.) we can write $\hat{\phi}(x)$ and
its time derivative $\hat{\pi}(x)$ as
\begin{subequations}
\begin{eqnarray}
\hat{\phi}(x)\stackrel{\text{H.a.}}{=}\phi(x) + \sum_{\alpha}
\left[ \hat{b}^\dagger_{\alpha} f_{\alpha}(x) +
\hat{b}_{\alpha} f_{\alpha}^*(x)\right], \\
\hat{\pi}(x)\stackrel{\text{H.a.}}{=}\pi(x) + \sum_{\alpha}
\left[ \hat{b}^\dagger_{\alpha} \dot{f}_{\alpha}(x) +
\hat{b}_{\alpha} \dot{f}_{\alpha}^*(x)\right],
\end{eqnarray}
\label{eq:hartree}
\end{subequations}
where $\phi(x)=\langle\hat\phi(x)\rangle$
and $f_\alpha(x)$ are the time-dependent mean field and
mode functions, and $\hat{b}_\alpha$ are time-\emph{in}{\-}de{\-}pen{\-}dent
creation and annihilation operators satisfying the usual commutation relations.
Without an approximation a similar expansion is possible but only with
time-dependent creation and annihilation operators:
in terms of the initial $\hat b_{\alpha}$ and $\hat b_{\alpha}^{\dagger}$
the dependence is then non-linear at later times. The fact that the same set
can be used at all times is due to the Gaussianity of the Hartree approximation
(see e.g.~\cite{SaSm00a} for more information). Combining
Eqs.~\eqref{eq:hartree} and \eqref{eq:opeq} and taking expectation values gives
the
equations of motion for $\phi$ and the $f_\alpha$,
\begin{subequations}
\begin{eqnarray}
   \ddot{\phi} & = &\Delta \phi_{\hphantom{\alpha}} - [\mu^2 +
   \hphantom{3}\lambda \phi^2 +
    3 \lambda C ] \phi, \label{eq:phieom} \\
   \ddot{f}_{\alpha} & = &\Delta f_{\alpha} - [\mu^2 + 3\lambda \phi^2
   + 3 \lambda C ] f_{\alpha}, \label{eq:modeeom}
\end{eqnarray}
with
\begin{equation}
C = \sum_{\alpha} (2n_\alpha^0 + 1)|f_{\alpha}|^2, \quad
n_\alpha^0 = \langle \hat{b}^\dagger_{\alpha} \hat{b}_{\alpha} \rangle.
\label{eq:modesum}
\end{equation}
\label{eq:eom}
\end{subequations}
The exact equation of motion for the one-point function, which
follows by taking the expectation value of the Heisenberg equation of
motion \eqref{eq:opeq}, contains the three-point function. An exact equation for
the three-point function can be obtained from taking the expectation value of
the product of Eq.~\eqref{eq:opeq} with 
two 
field operators and will contain the
five-point function. The resulting infinite hierarchy of higher point functions
is truncated by the Hartree approximation which factorizes these
higher-point functions into one- and two-point functions.

\subsection{Choice of mode functions and renormalization
\label{subsec:inimodes}}

As in our previous work, we will use the stationary solutions of
Eqs.~\eqref{eq:eom} to define the initial mode functions, i.e.~we will
choose them as plane waves with wave vector $k$ ($\alpha \rightarrow k$)
\begin{equation}
  f_k(x,0)=\frac{e^{ikx}}{\sqrt{2\omlat{0} L}}, \qquad
  \dot{f}_k(x,0) = -i\omlat{0} \frac{e^{ikx}}{\sqrt{2\omlat{0} L}}.
\label{eq:inimode}
\end{equation}
Here $L$ is the ``volume'' of our one dimensional system (using periodic
boundary conditions) and $\omlat{0} = \sqrt{m^2 + k^2}$, with
$m$ the physical mass, defined self-consistently as the square root of the term
in square brackets in Eq.~\eqref{eq:modeeom}.
The sum \eqref{eq:modesum} for these initial mode functions diverges
logarithmically in 1+1 D, which is compensated by the bare $\mu^2$.
In the vacuum, i.e.~using stationary solutions for $\phi$ and $f_\alpha$ of
Eqs.~\eqref{eq:eom} with $n_\alpha^0=0$, the renormalized mass parameter
and mode sum are given by
\begin{equation}
\mu_{\text{ren}}^2 = \mu^2 + 3 \lambda C_{n_{\alpha}^0=0},
\quad
\mu^2 + 3\lm C = \mu_{\text{ren}}^2 + 3\lm C_{\text{ren}}.
\label{eq:massrenorm}
\end{equation}
Then
$m^2=\mu_{\text{ren}}^2 + 3\lambda v^2$, with $v$ the stationary value of
$\phi$.

In this paper we will use $n_\alpha^0 = 0$.
Using the mode functions \eqref{eq:inimode},
the coherent states $|\ph,\pi\rangle$ satisfying $b_{k} |\ph,\pi\rangle = 0$,
have mean fields $\ph(x)$ and $\pi(x)$, with a variance that is easily
computed.
We shall use these states as basis for our Hartree ensembles \eqref{eq:HEA}.

\subsection{Initial conditions \label{subsec:inimf}}

We used two different initial conditions for the mean field, a sum of standing
waves with a flat distribution of phases, which we also
used in our previous work in the \brokenph{} \cite{SaSm00a}, and a single
Gaussian wave packet, as studied by Bettencourt et al.~\cite{BePa01}.

The first is given by
\begin{equation}
\phi(x) = 0, \qquad \pi(x) = A m \sum_{j=1}^{j_{\text{max}}} \cos(2 \pi j x/L -
\psi_j),
\label{eq:iniwaves}
\end{equation}
where
the maximum momentum
$2 \pi j_{\text{max}}/L$ is typically of the order of the mass $m$ and
the constants $\psi_j$ are random phases with a flat distribution
(i.e.~they are uniformly distributed in $[0,2\pi)$).
We shall call such $\rh[\ph,\pi]$ flat ensembles.
The energy, which is independent of the phases $\psi_j$, is given by
\begin{equation}
\frac{E}{m} = \frac{A^2 L m j_{\text{max}}}{4}.
\end{equation}
We use both $A$ and $j_{\text{max}}$ to vary the total energy density.

The second initial condition is a Gaussian wave packet:
\begin{equation}
\pi(x) = 0, \qquad \phi(x)=\Phi \exp\left[-\frac{x^2}{2 A}\right].
\label{eq:inigauss}
\end{equation}
Its energy is given by
\begin{equation}
\frac{E}{m} = \frac{\Phi^2}{8} \sqrt{\frac{\pi}{Am^2}} \bigl( 2 + 4 A
m^2 + \sqrt{2} A \lambda \Phi^2 \bigr).
\end{equation}
We will restrict ourselves to $A m^2=2$ and use $\Phi$ to vary the total energy
in the system. For this type of initial conditions we do not average over
multiple runs, so $\rh[\ph,\pi]$ is a delta functional and $\hat\rh$
is a coherent pure state.

\subsection{Observables \label{subsec:observ}}

As in \cite{SaSm00a} we define particle numbers $n_k$ and frequencies $\om_k$
in terms of equal time correlation functions,
\begin{subequations}
\begin{align}
S(x,y) &= \overline{\langle \hat{\phi}(x) \hat{\phi}(y) \rangle} -
\overline{\langle \hat{\phi}(x) \rangle}\;\;
\overline{ \langle \hat{\phi}(y) \rangle},\\
U(x,y) &= \overline{\langle \hat{\pi}(x) \hat{\pi}(y) \rangle} -
\overline{\langle \hat{\pi}(x) \rangle}\;\;
\overline{\langle \hat{\pi}(y) \rangle}.
\end{align}
\label{eq:2pnt}
\end{subequations}
\noindent
The over-line indicates coarse graining over all space and, depending on the
simulation, also includes course graining over a time interval
and an ensemble of initial conditions.
The symmetrized $\pi \phi$-correlation function, $T(x,y)$,
vanishes in equilibrium, and we will just use $S$ and $U$.
Taking the average over all of space leaves $S$ and $U$ depending only on the
coordinate difference $x-y$. In terms of the Fourier transform,
\begin{equation}
S_k = \int dx\, e^{-ikx}\, S(x,0)
\end{equation}
and similar for $U$, the time-dependent
particle numbers and frequencies are defined by
\begin{equation}
S_k = \left(n_k + \frac{1}{2}\right) \frac{1}{\om_k}, \qquad
U_k = \left(n_k + \frac{1}{2}\right) \om_k.
\label{eq:twopoint}
\end{equation}
Therefore, apart from numerical corrections discussed in
Sec.~\ref{app:latomega},
we define instantaneous particle number and
frequency by
\begin{equation}
n_k = \sqrt{U_k S_k} - \frac{1}{2}, \qquad \omega_k = \sqrt{\frac{U_k}{S_k}}.
\label{eq:quasipart}
\end{equation}
In practice $n_k$ is positive (it can be shown to
be positive provided the symmetric correlation between $\ph$ and $\pi$
vanishes).
Other definitions of particle numbers are in use, e.g.\ derived  
from time-dependent creation and annihilation operators defined in terms of
adiabatic mode functions, 
\begin{multline}
\hat{a}_k(t)\, e^{-i\int_0^t dt'\, \tilde\omega_k(t')} = \\
\frac{1}{\sqrt{2\tilde\omega_k(t) L}}
\int_0^L dx\, e^{-ikx}\, [\tilde\omega_k(t)\hat\phi(t,x) + i \hat\pi(t,x)].
\end{multline}
However, one then still has to choose $\tilde\omega_k(t)$.  
This can introduce some ambiguity, especially if the system is far from 
equilibrium and the effective mass term in the equations of motion 
for the modes becomes negative; see for example 
\cite{CoHa97,BoVe96}. Using Eq.~\eqref{eq:quasipart} 
as the definition of particle number, has the advantage
that the frequencies are real and positive by construction.

The correlation functions $S$ and $U$ may be written as a sum of contributions
from the mean field and  mode functions separately. Given $\om_k$ as
obtained from the total $S$ and $U$,
we can study their separate contributions to $n_k$ as
defined by Eqs.~\eqref{eq:twopoint}.

The energy density can be obtained from the effective Hamiltonian
$H_{\text{eff}} = \langle \hat H\rangle$ \cite{SaSm00a}.
We split it into mean field and mode contributions according to
\begin{subequations}
\label{eq:effhamil}
\begin{align}
E_{\text{mf}} = &
\frac{1}{2} (\partial_t \phi)^2 + \frac{1}{2} (\partial_x \phi)^2
+ \frac{1}{2} \mu_{\text{ren}}^2 \phi^2 + \frac{1}{4} \lambda \phi^4
\nonumber\\
& - \frac{1}{2}\mu_{\text{ren}}^2 v^2 - \frac{1}{4} \lm v^4,\\
E_{\text{modes}} =&
\sum_\alpha \left(\frac{1}{2} + n_\alpha^0\right)
\bigl( |\partial_t f_\alpha|^2 + |\partial_x f_\alpha|^2 \bigr)
+ \frac{1}{2} \mu_{\text{ren}}^2 C_{\text{ren}}
\nonumber \\
&
+ \frac{3}{2} \lambda C_{\text{ren}} \phi^2
+ \frac{3}{4} \lambda C_{\text{ren}}^2
- H_{\text{vac}}
\end{align}
\end{subequations}
where $v$, the ground-state value of $\phi$,
is zero in the \symmetricph{}.
Although the splitting between modes and mean field is somewhat
arbitrary, the above definition has the advantage that it reduces to the
classical expression when the modes are of the vacuum form \eqref{eq:inimode}.

The quasi-particle energy is also an interesting observable. It
may be defined as
\begin{equation}
E_{\text{qp}} = \sum_k n_k \omega_k \label{eq:partener}
\end{equation}
where $n_k$ can be obtained from the two-point functions of the mean field, of
the mode functions, or of the sum of both.

\subsection{Implementation on a lattice \label{app:latomega}}

The discretization of the scalar field theory on a space-time lattice
has some elegant features
which we present briefly in this section; for fermions, see \cite{AaSm99}.
For simplicity we start with a simple quantum mechanical system of unit mass,
with action
\begin{equation}
S=a_0\sum_t \left\{\frac{[q(t+a_0)-q(t)]^2}{2 a_0^2} - V(q(t))\right\},
\end{equation}
where $a_0$ is the time step, $t=a_0 r$, with integer $r$.
We define the quantum system by means of the path integral.
The discretized path integral
\begin{equation}
Z=\int\Bigl[\prod_t dq(t)\Bigr] e^{iS}
\end{equation}
corresponds to
an evolution operator in Hilbert space that is a product of single step
evolution operators given by
\begin{equation}
\hat U= \hat U_p \hat U_q,
\end{equation}
with
\begin{equation}
\hat U_p = e^{-ia_0 \hat p^2/2},
\quad
\hat U_q = e^{-ia_0 V(\hat q)},
\end{equation}
where $\hat p$ and $\hat q$ are canonical operators satisfying
$[\hat q,\hat p ]=i$.
A finite time evolution then takes the ``Trotter form''
\begin{equation}
\hat U_q \hat U^r = \hat U_q
\cdots \hat U_p \hat U_q \hat U_p \hat U_q \hat U_p\hat U_q
\cdots \hat U_q,
\end{equation}
The Heisenberg operators
\begin{equation}
\hat p(t) = \hat U^{r\dagger} \hat p\, \hat U^r,
\quad
\hat q(t) = \hat U^{r\dagger} \hat q\,  \hat U^r,
\quad
t=a_0 r,
\end{equation}
satisfy the discretized equations of motion in leapfrog fashion,
\begin{subequations}
\label{leapfqp}
\begin{eqnarray}
\hat p(t+a_0) &=& \hat p(t) - a_0 V'(\hat q(t)),\\
\hat q(t+a_0) &=& \hat q(t) + a_0 \hat p(t+a_0).
\label{leapfqpb}
\end{eqnarray}
\end{subequations}
With $\hat q(t) \to q(t)$, $\hat p(t) \to [q(t)-q(t-a_0)]/a_0$,
the above Eqs.~\eqref{leapfqp} are identical in form to the
classical equations obtained from the stationary action principle.

Making a unitary transformation
\begin{equation}
\hat T = e^{-ia_0 V(\hat q)/2}\, \hat U e^{ia_0 V(\hat q)/2},
\end{equation}
we get an equivalent operator $\hat T$,
which becomes the Hermitian and positive transfer operator
upon analytically continuing to imaginary time (see e.g.~\cite{Sm02}),
writing $a_0 = e^{-i\theta}|a_0|$, $\theta = 0 \to \pi/2$,
\begin{equation}
\hat T \to
e^{-|a_0| V(\hat q)/2}\, e^{-|a_0|\hat p^2/2}\, e^{-|a_0| V(\hat q)/2}.
\end{equation}

Specializing to the harmonic case $V(q) = \om^2 q^2/2$ we can diagonalize the
time evolution in terms of creation and annihilation operators
$\hat c^{\dagger}$ and $\hat c$,
\begin{equation}
\hat T \hat c \hat T^{\dagger} = e^{ia_0\om^{(e)}}\, \hat c,
\quad
\hat T \hat c^{\dagger} \hat T^{\dagger} = e^{-ia_0\om^{(e)}}\,
\hat c^{\dagger},
\end{equation}
with
\begin{equation}
\hat c = \frac{1}{\sqrt{2\om^{(n)}}}(\om^{(n)} \hat q + i\hat p)
\end{equation}
and
\begin{subequations}
\begin{align}
\cos(a_0\om^{(e)}) &= 1 - \frac{1}{2} a_0^2 \om^2,
\label{defome}\\
\om^{(n)} &= \frac{1}{a_0}\sin(a_0\om^{(e)})
= \omega \sqrt{1-\frac{1}{4} a_0^2 \omega^2},
\label{defomn}
\end{align}
\end{subequations}
and the conjugate relation for $\hat c^{\dagger}$.
The creation and annihilation operators satisfy the standard commutation
relation $[\hat c,\hat c^{\dagger}] = 1$. The superscripts $e$ and $n$
distinguish the ``exponent omega'' (eigenvalue omega) $\om^{(e)}$
from the ``normalization omega'' (eigenvector omega) $\om^{(n)}$, and both
go over to the ``original omega'' $\om$ in the continuous time limit $a_0\to 0$.
The ground state is given by
\begin{equation}
\hat c |0\rangle = 0,
\quad
\langle q|0\rangle = \nu e^{-\om^{(n)} q^2/2},
\end{equation}
with $\nu$ a normalization constant and
\begin{equation}
\hat T  (\hat c^{\dagger})^n |0\rangle = e^{-i(n+1/2)a_0\om^{(e)}}\, 
(\hat c^{\dagger})^n |0\rangle .
\end{equation}
The evolution becomes unstable when $a_0^2\om^2 > 4$, for which $\omlat{e}$
is imaginary. The eigenvalues of $\hat T$ are then no longer phase factors
and its eigenfunctions no longer normalizable, despite its formally
unitary form. This is of course avoided by taking $a_0$ sufficiently small.
The discretization errors in $\om^{(e)}$ and $\om^{(n)}$ are of order
$a_0^2$.

It is natural to identify the Hamiltonian $\hat H$ from
$\hat T =\exp(-ia_0 \hat H)$,
but this leaves a modulo $2\pi/a_0$ ambiguity for the eigenvalues of
$\hat H$ (the imaginary time version is unambiguous). To pin down $\hat H$
more precisely we can use the Baker-Campbell-Hausdorff series for combining
the exponents in $\hat T$, which gives $\hat H = \hat p^2/2 + V(\hat q)
+ {\cal O}(a_0^2)$. We shall neglect the corrections of order $a_0^2$.
The exact $\hat H$ is time independent. In practice, the expectation
value of the approximate $\hat H$ is constant in time up to small fluctuations,
as expected for a leapfrog algorithm.

For application to the Hartree approximation
it will be more convenient for us to work with the unitarily related
creation and annihilation operators that diagonalize $\hat U$,
\begin{eqnarray}
\hat a &=& e^{ia_0 V(\hat q)/2}\, \hat c\, e^{-ia_0 V(\hat q)/2}
\nonumber\\
&=& \frac{1}{\sqrt{2\om^{(n)}}}\left(\frac{1-e^{-ia_0\om^{(e)}}}{ia_0}\hat q
+ i\hat p\right),
\label{moreconv}\\
\hat U \hat a \hat U^{\dagger} &=& e^{ia_0\om^{(e)}}\, \hat a,
\end{eqnarray}
for $V = \om^2 q^2/2$.
Note that $\hat{a} \to \hat{c}$ in the limit $a_0 \to 0$.

The generalization of the above quantum mechanical model to our scalar field
is straightforward. The lattice action on a space-time lattice
with spatial-temporal lattice distance $a/a_0$
is given by
\begin{equation}
\begin{split}
S[\phi]= a_0 a\sum_{x,t} \biggl\{ &
\frac{\left[\phi(x,t+a_0)-\phi(x,t)\right]^2}{2 a_0^2}
\\
& - \frac{\left[\phi(x+a,t)-\phi(x,t)\right]^2}{2 a^2} -
\frac{1}{2} \mu^2\phi(x,t)^2
\\
& - \frac{1}{4} \lambda\phi(x,t)^4\biggr\},
\end{split}
\end{equation}
where we assume a periodic physical size $L = Na$.
The operator description in Hilbert space
follows from the lattice regularized path integral.
In the Hartree approximation we write the operator fields in terms of a
complete set of mode functions,
\begin{subequations}
\begin{align}
\hat \phi(x,t) &= \phi(x,t)
+ \sum_k [\hat b_k f_k(x,t) + \hat b_k^{\dagger} f_k(x,t)^*],\\
\hat\pi(x,t) &= \pi(x,t)
+ \sum_k [\hat b_k \dot f_k(x,t)
+ \hat b_k^{\dagger} \dot f_k(x,t)^*],
\end{align}
\end{subequations}
where the use of
\begin{equation}
\dot f(x,t) = \frac{f(x,t) - f(x,t-a_0)}{a_0}
\end{equation}
is inspired by Eq.~\eqref{leapfqpb}.
(Using instead the forward derivative
$\dot f_k(x,t) = [f(x,t+ a_0) - f(x,t)]/a_0$ gives equivalent results.)
Imposing canonical commutation relations for both
$\hat{\phi}$, $\hat{\pi}$ and $\hat{b}_k$, $\hat{b}^\dagger_k$,
leads to the orthonormality and completeness relations
\begin{subequations}
\begin{align}
a\sum_x [i\dot{f}_k(x,t) f^*_l(x,t) -if_k(x,t) \dot{f}^*_l(x,t)] &=
\delta_{kl}, \label{eq:ortho} \\
\sum_k [if_k^*(x,t) \dot{f}_k(y,t) -i f_k(x,t) \dot{f}^*_k(y,t)] &=
\frac{\delta_{xy}}{a}.
\label{eq:compl}
\end{align}
\end{subequations}
The time independence of the orthonormality conditions corresponds to
Noether charges of symmetries \cite{SaSm00a} of the effective action on
the lattice.
We use the static solutions of the Hartree equations
in constructing the set of mode functions. Their equation of motion
\begin{multline}
\label{eq:eomlat}
\frac{f_k(x,t+a_0) -2 f_k(x,t) + f_k(x,t-a_0)}{a_0^2} = \\
\frac{f_k(x+a,t) - 2 f_k(x,t) + f_k(x-a,t)}{a^2}
- m^2 f_k(x,t),
\end{multline}
can be written in the leapfrog form \eqref{leapfqp}.
The solution of the recursion relation \eqref{eq:eomlat} can be written as
\begin{equation}
f_k(x,t)=\frac{e^{i k x -i \omlat{e} t}}{\sqrt{2 \omlat{n} L}},
\quad
k=\frac{2\pi j}{L},
\quad
j=-\frac{N}{2} + 1, \cdots, \frac{N}{2},
\label{eq:inimodelat}
\end{equation}
giving
\begin{equation}
- \frac{2-2 \cos(\omlat{e} a_0)}{a_0^2} + \frac{2-2 \cos(k a)}{a^2} + m^2 = 0.
\label{eq:omexp}
\end{equation}
Defining a lattice $\omlat{a}$ as
\begin{equation}
\omlat{a} = \sqrt{m^2 + \frac{2-2 \cos(k a)}{a^2}},
\label{eq:defomcont}
\end{equation}
we find the analogue of Eq.~\eqref{defome},
\begin{equation}
\cos(a_0\omlat{e}) = 1 - \frac{1}{2} a_0^2 (\omlat{a})^2,
\label{eq:defomexp}
\end{equation}
which has real $\omlat{e}$ solutions for $a/a_0 > \sqrt{4 + a^2 m^2}$.
We used $a/a_0 = 10$ in our simulations, which amply secured stability.
The normalization in Eq.~\eqref{eq:inimodelat} is fixed by the
orthonormality relation \eqref{eq:ortho}, which gives the analogue of
Eq.~\eqref{defomn}
\begin{equation}
\omlat{n} = \frac{\sin(a_0 \omlat{e})}{a_0} =
\omlat{a} \sqrt{1-\frac{1}{4} a_0^2 (\omlat{a})^2},
\label{eq:defomnorm}
\end{equation}
The completeness relation \eqref{eq:compl} is then also satisfied.
When the mode functions have the form \protect \eqref{eq:inimodelat},
the $\hat a_k$ defined by $\hat\phi = \sum_k \hat a_k f_k + \text{h.c.}$,
$\hat\pi = \sum_k \hat a_k \dot f_k + \text{h.c.}$,
are related to $\hat\phi$ and $\hat\pi$ as in the quantum mechanical case
\eqref{moreconv}.
Note that $\omlat{n}, \omlat{e} \to \omlat{a}$ in the limit $a_0 \to 0$,
and $\omlat{a} \to \sqrt{m^2 + k^2}$ as $a\to 0$.

We end this section with a properly discretized version of the
instantaneous particle number $n_k$, using the stationary solution
\eqref{eq:inimodelat} and the two-point functions \eqref{eq:2pnt}.
Suppose the mean field is zero and
\begin{equation}
\langle\hat b_k^{\dagger}\hat b_k \rangle = n_k^0 = n_{-k}^0.
\end{equation}
Then
\begin{subequations}
\begin{align}
S_k(t) &= \left(n_k^0 + \frac{1}{2}\right)\frac{1}{\omlat{n}}, \\
U_k(t) &= \left(n_k^0+\frac{1}{2}\right) \frac{(\omlat{a})^2}{\omlat{n}},
\end{align} \label{eq:twopointlatt}
\end{subequations}
where we used
\begin{equation}
\dot{f}_k(x,t)\dot f_k^*(y,t) = (\omlat{a})^2 f_k(x,t) f_k^*(y,t).
\end{equation}
Inverting Eqs.~\eqref{eq:twopointlatt} we find that our definition
of the instantaneous energy $\omega_k(t)$ does not need discretization
corrections,
\begin{equation}
\omlat{a} = \sqrt{\frac{U_k(t)}{S_k(t)}} \equiv
\omega_k(t).
\label{eq:defominst}
\end{equation}
On the other hand, the definition of instantaneous particle number
needs important corrections for large $\om_k$:
\begin{equation}
n_k^0+\frac{1}{2}
= \sqrt{U_k S_k} \frac{\omlat{n}}{\omlat{a}}
= \sqrt{U_k ( S_k - \frac{1}{4} a_0^2 U_k)}
\equiv n_k(t) + \frac{1}{2},
\label{eq:nlat}
\end{equation}
using Eqs.~\eqref{eq:defomnorm} and \eqref{eq:defominst}.

For larger energies the corrections can become quite important.
Denoting the uncorrected particle number by
$\tilde{n}_k=\sqrt{U_k S_k}-1/2$, we find
\begin{equation}
\begin{split}
\frac{\tilde{n}_k - n_k}{n_k} &= \frac{n_k+\frac{1}{2}}{n_k}
\Bigl( \frac{1}{\sqrt{1-\frac{1}{4} (a_0 \omlat{a})^2}} - 1 \Bigr) \\
& \approx \frac{n_k+\frac{1}{2}}{n_k} \frac{1}{8} (a_0 \omlat{a})^2.
\end{split}
\end{equation}
Using a Bose-Einstein distribution at $T=m$ and the typical value $a_0 m = 1/80$
we find that the relative difference becomes unity for $\omega_k/m = 7.5$.
At the lower temperature $T/m=0.5$ this is the case already for
$\omega_k/m = 4.3$.
\section{Numerical results \label{sec:numres}}

In this section we will present data we obtained using numerical simulations in
the \symmetricph{},
i.e.~$\mu^2$ and $\lambda$
corresponding to the \symmetricph{} at zero temperature.
First we will discuss the particle distribution to study its equilibration
behaviour and to search for thermalization. Then we will examine the energies
and auto-correlation function to analyse the time scales in the theory.

\subsection{Flat ensemble \label{subsec:highener}}

In the flat ensemble of initial conditions,
the initial mean field $\phi$ of a realization is equal to its
vacuum expectation value $0$, while its momentum is the sum of waves
with random phase, as specified in
Eqs.~\eqref{eq:iniwaves}.
\begin{figure}[tbp]
\includegraphics[width=0.47\textwidth]{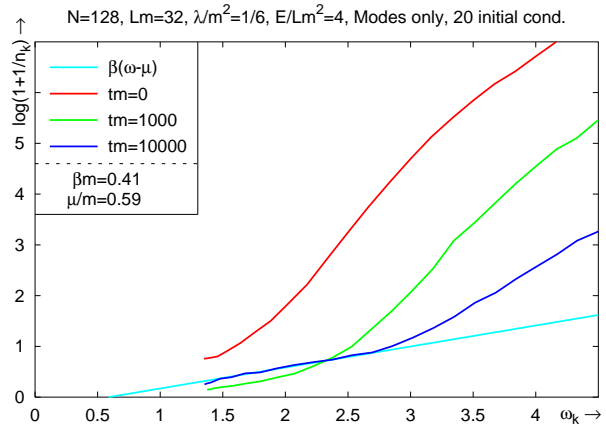}
\caption{\label{fig:dist_flat_early} The particle numbers
$\log(1+1/n_k)$ in the modes
as a function of $\omega_k$. Average over 20 flat ensemble initial conditions,
$\lambda/m^2=1/6$, $E/Lm^2=4$, at times up to $tm=$ $10^4$. Time increases from
the top curve to the bottom curve.}
\end{figure}
We averaged over $10$ or $20$ initial
conditions, and excited all non-zero modes up to
$k_{\text{max}}= 2\pi j_{\text{max}}/L = \pi m/2$.
Simulations have
been carried out for three different couplings $\lambda/m^2=1/6, 1/8, 1/12$ and
three different energy densities $E/Lm^2=4, 2, 1$, 
as well as for the combination $\lambda/m^2 = 0.1$ and $E/Lm^2 = 0.4$.
We mainly used $N=128$ lattice points, and volume $Lm=32$, while the temporal
lattice distance $a_0=a/10$.

\subsubsection{Particle distribution function \label{subsubsec:dist_highener}}

Fig.~\ref{fig:dist_flat_early} shows the particle number obtained for
coupling $\lambda/m^2=1/6$ and energy density $E/Lm^2=4$.
As in our previous work we compare the
out-of-equilibrium particle densities with a Bose-Einstein (BE) distribution
\begin{equation}
n_k = \frac{1}{e^{(\omega_k - \mu)/T}-1}.
\label{eq:be}
\end{equation}
We therefore plotted $\log(1+1/n_k)$ versus $\omega_k/m$, since a
BE distribution then shows up as a straight line with slope $m/T$ and offset
$-\mu/m$ ($T$ temperature, $\mu$ chemical potential).
For this largest coupling and energy density in our study
we find approximate thermalization to the BE form
with a temperature $T/m=2.4$ and chemical
potential $\mu/m=0.6$. In contrast to what was found in the \brokenph{}
\cite{SaSm00a,SaSm00b},
a substantial chemical potential is needed to make a reasonable fit.
Another difference is the larger time scale involved:
in the \brokenph{} at an energy density $E/Lm^2=0.5$ and the same
$\lambda/|\mu_{\text{ren}}^2| = 1/6$ ($\lambda/m^2=1/12$),
we could already recognize BE behaviour with $T/m \approx 1$
at a time $tm\lesssim 100$, while here, at an $8$
times larger energy and roughly $2$ times larger
effective BE temperature we can only clearly do so at time $tm\gtrsim 2000$.
A fit of the local temperature $T(t)$ approaching approximate equilibrium
gives an equilibration-time scale $m\tau_{\text{BE}}=1500-1600$ (exponential
fit over $k/m<1.7$, $100<t<6000$).

\begin{table}[tbp]
\begin{tabular}{lc|r@{.}l|r@{.}l|r@{.}l|}
& & \multicolumn{2}{l|}{$E/Lm^2=1$} & \multicolumn{2}{c|}{$E/Lm^2=2$} &
\multicolumn{2}{r|}{$E/Lm^2=4$} \\
\hline
$\lambda/m^2=1/6$ & $\beta m$ & $1$&$12$ & $0$&$71$ & $0$&$41$ \\
                  & $\mu/m$ &   $0$&$95$ & $0$&$83$ & $0$&$59$ \\
\hline
$\lambda/m^2=1/8$ & $\beta m$ & $0$&$89^*$ & $0$&$60$ & $0$&$45$ \\
                  & $\mu/m$ &   $0$&$62$ & $0$&$76$ & $0$&$80$ \\
\hline
$\lambda/m^2=1/12$ & $\beta m$& \multicolumn{2}{l|}{$--$} & $0$&$68^*$ &
$0$&$40$ \\
           & $\mu/m$ &  \multicolumn{2}{l|}{$--$} & $0$&$76$ & $0$&$85$
\\
\hline
\end{tabular}
\caption{\label{tab:flat_temp} Inverse temperature $\beta$ and chemical
potential $\mu$ as derived from a Bose-Einstein fit to the particle numbers
(modes only). See text for further explanation.}
\end{table}
For most parameters used in our simulations we can recognize
Bose-Einstein features in the low momentum part of the
distributions, and linear fits can be made 
as in Fig.~\ref{fig:dist_flat_early}.
The results of these fits are shown in Table~\ref{tab:flat_temp}.
Fits marked with a star were made at
$tm=$ $59000\cdots 60000$, all others at
$tm=9000\cdots 10000$. Including
the mean field in the two-point functions,
typically gives the same temperature within errors,
but a noticeably larger value (+ 0.05) for $\mu/m$,
corresponding to a higher particle number.
The chemical potential is also more sensitive to
the exact fit-interval than the temperature.
In the two *-marked runs a thermal
distribution could be recognized
only at times
$\gtrsim 20000$ ($\lambda/m^2=1/12$)
and $\gtrsim 45000$ ($\lambda/m^2=1/8$).

For the run at $\lambda/m^2=1/12$, $E/Lm^2=1$, we
did not find a thermal-like distribution even
at the latest simulation time $tm = 10^5$.
We see that the energy is transferred from the mean
field to the modes and the system equilibrates ``locally in $k$'',
but the total particle number remains roughly unchanged.
The same was found at the lower energy density $E/Lm^2=0.4$, and also
for the Gaussian wave packet initial condition.
We interpret this as a
resonance phenomenon in the equation of motion of the mode
functions, which will be described in Section~\ref{subsec:localtherm}.

\begin{figure}[tbp]
\includegraphics[width=0.48\textwidth]{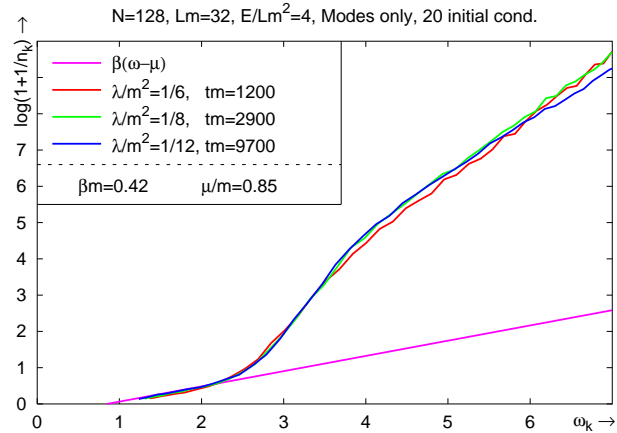}
\caption{\label{fig:dist_flat_compar}
Scaling behaviour of the particle distribution at fixed energy
density and differing couplings and times.}
\end{figure}
\begin{figure}[tbp]
\includegraphics[width=0.48\textwidth]{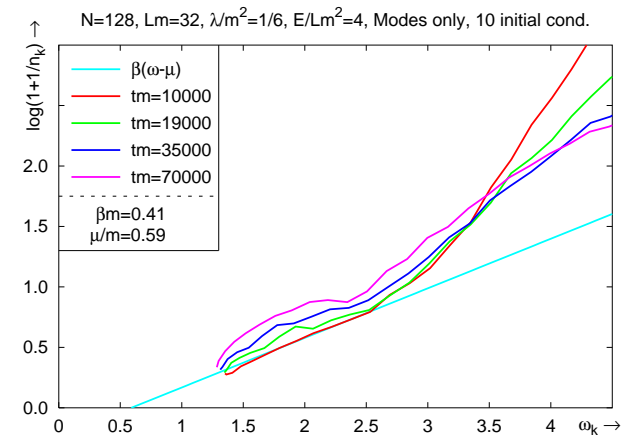}
\caption{\label{fig:dist_flat_late} The particle numbers
$\log(1+1/n_k)$ in the modes as a function of $\omega_k$
at large times $tm>10^4$. In the small $k$ region time increases from bottom
to top, in the large $k$ region time increases from top to bottom.}
\end{figure}
Comparing the results at $E/Lm^2=2$ and $4$ it seems that the temperature
only depends on the energy density and not on the coupling.
This appears to hold even for the distribution function itself,
as illustrated in
Fig.~\ref{fig:dist_flat_compar}, where the distributions for different
couplings are plotted at different times.
The different times at which the curves in the figure overlap
suggest that the
equilibration time scale for the particle distribution
is proportional to $\lambda^{-3}$. The same power is found at
the energy density $E/Lm^2=2$.
Table~\ref{tab:flat_temp} shows that the temperature is roughly
proportional to $\sqrt{E/L}$, which can be understood from the scaling
behaviour in
Fig.~\ref{fig:dist_flat_compar}, since there is no other scale left.
The same argument should apply to the chemical potential. However, this
quantity is more dependent on time than the temperature and runs at
different parameters are best compared at different times as
in Fig.~\ref{fig:dist_flat_compar},
which we have not done in Table~\ref{tab:flat_temp}.

The independence of coupling
suggests that a representation of the energy
in terms of almost free quasi-particles will be reasonably good.
We will check this in the next section.

Fig.~\ref{fig:dist_flat_late} shows the distribution 
for late times, when it starts to deviate from
the BE form. Note the difference in vertical scale compared to
Fig.~\ref{fig:dist_flat_early}.
At $tm=15000 - 20000$
classical-like deviations become visible
in the form of concave behaviour at low $\om_k$ ($n_k = T/\om_k \Rightarrow
(\partial/\partial\om_k)^2 \log(1+1/n_k) < 0$).

\begin{figure}[tbp]
\begin{center}
\includegraphics[width=0.48\textwidth]{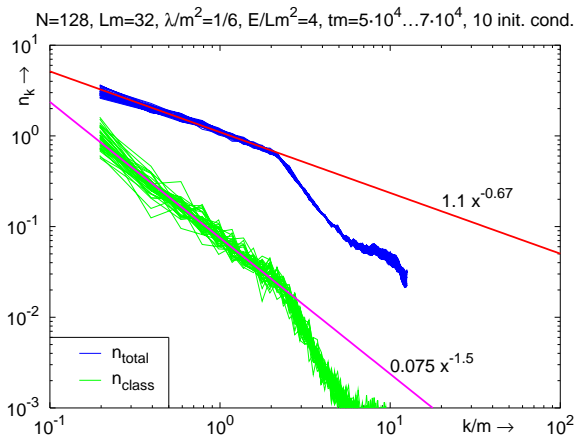}
\end{center}
\caption{Log-log plot of the particle numbers for mean field alone, and combined
with modes, vs $k/m$, at large times.\label{fig:powerlaw}}
\end{figure}

The mean field in this time region behaves very interestingly. In
Fig.~\ref{fig:powerlaw} we plotted the particle numbers at
$tm = 50000 \cdots 70000$ as a function of
momentum $k$, both for the mean field alone and for the total two-point
function, using a log-log scale  and
leaving out the zero mode. While the high-momentum modes
are still exponentially suppressed, the low-momentum modes have acquired a
power law distribution. The quantum-modes-only distribution does
{\em not} behave as a power law (cf.~Fig.~\ref{fig:dist_flat_late}).
The particle numbers as obtained from the mean field only and those
including the modes have different powers, $-1.5$ and $-0.67$ respectively.
Already much earlier, around $tm=8000$, this distribution starts to emerge, with
20\% larger powers.

The power-law behaviour in the low momentum modes of the mean field
apparently influences the quantum modes, in that their low
momentum modes are enhanced in comparison to the classical $T/\omega_k$.
We have seen this clearly in a plot of $n_k \omega_k$ ($\to T$ for
classical thermal equilibrium) showing a peak at $k=0$
and a ``classical plateau'' at the interval $k/m=1.0\dots2.2$.
Similar behaviour has also been found in the other runs at
$\lambda/m^2=1/6$, $E/Lm^2=2$ and $\lambda/m^2=1/8$, $E/Lm^2=4$.

In a purely classical simulation using the same set of parameters power-law
behaviour is {\em not} found.
This suggests that the interaction of the mean field with
with the quantum modes plays a crucial role,
even though the latter do not show power-law behaviour.
\begin{figure}[tbp]
\includegraphics[width=0.48\textwidth]{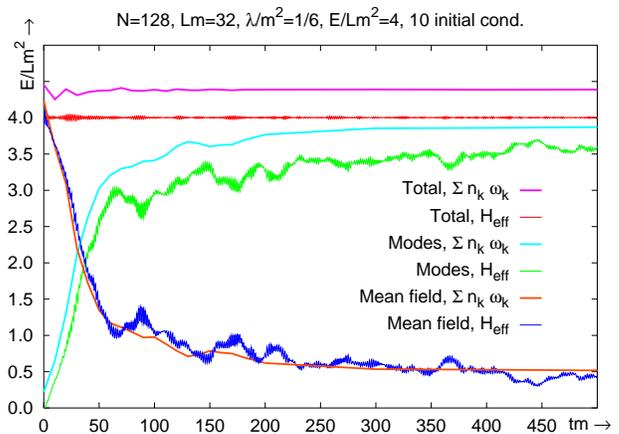}
\caption{\label{fig:ener_flat_short}
Contributions to the energy density, for short times.
>From top to bottom at $tm=450$:
$\sum_k n_k \om_k$ (mean field + modes), $E$ from $H_{\text{eff}}$,
$\sum_k n_k \om_k$ (modes only),
$E_{\text{modes}}$, $\sum_k n_k \om_k$ (mean field), $E_{\text{mf}}$.}
\end{figure}
\begin{figure}[tbp]
\includegraphics[width=0.48\textwidth]{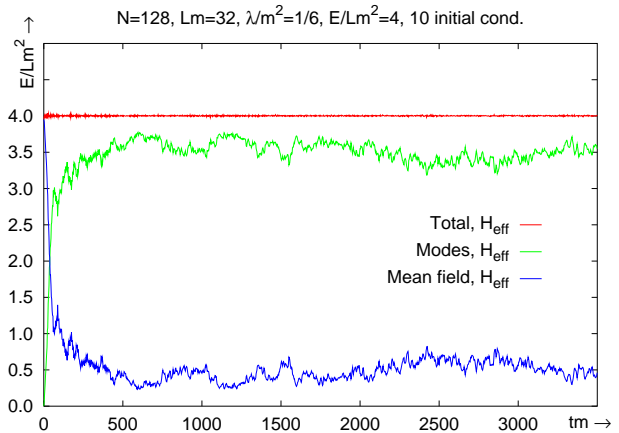}
\caption{\label{fig:ener_flat_medium}
Energy contributions for the same run as in Fig.~\ref{fig:ener_flat_short}
for intermediate times.}
\end{figure}
\begin{figure}[tbp]
\includegraphics[width=0.48\textwidth]{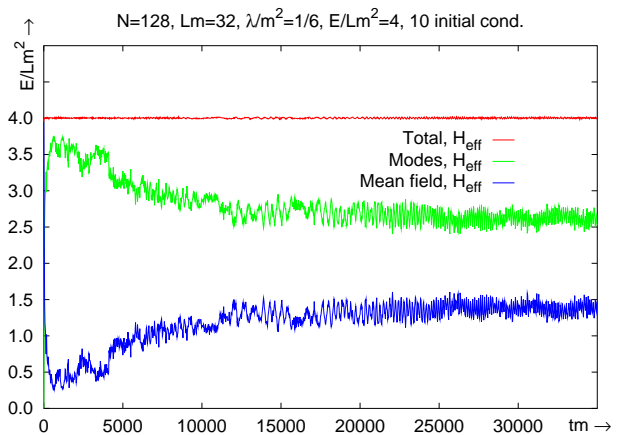}
\caption{\label{fig:ener_flat_long}
As in Fig.~\ref{fig:ener_flat_medium} for long times.}
\end{figure}

\subsubsection{Time scales for energy exchange\label{subsubsec:timesc_highener}}

In the previous section we obtained the scaling behaviour
$\propto \lm^{-3}$ for the time scale of approximate equilibration
based on the particle distribution. In this section
we will use the energy density in the different parts of the field to find the
short-time equilibration behaviour.
Figs.~\ref{fig:ener_flat_short}-\ref{fig:ener_flat_long} show the results
from one of the simulations
at $\lambda/m^2=1/6$, $E/Lm^2=4$, plotted at different time scales.
In Fig.~\ref{fig:ener_flat_short}, showing the early stage, the
energy as obtained from the quasi-particles is also included.
The quasi-particle picture appears to give a reasonable representation of
the energies, with a roughly constant $10\%$ mismatch in the total energy,
to which we will come back, below.
We furthermore see that the total energy in the quasi-particle picture is almost
constant, corresponding to a quasi-particle number that is itself
almost constant.
This is consistent with the chemical potential found in the BE fits.
We have checked that the dispersion relation of the quasi-particle energies is
close to that of free particles, $\om_k = \sqrt{m_{\text{eff}}^2 + k^2}$,
but with an effective mass $m_{\text{eff}}$ that is larger than
$m$, as can be seen, for example, from the minimum values of $\om_k$ in
Fig.~\ref{fig:dist_flat_early}, 
consistent with what follows from the effective potential.
Coming back to the $10\%$ mismatch in the quasi-particle energy: we have
checked that it is neither a finite volume nor a finite lattice distance
(spatial or temporal) effect. However, it does depend on the type of initial
conditions used. For example a simulation with Bose-Einstein type initial
conditions, as used in Ref.~\cite{SaSm00b}, with the same total energy, leads to
a much smaller mismatch of almost $3\%$. It is important to note that we are
looking at an effective description of an interacting theory and it is not
expected that the total energy in these particles is completely equal to the
actual total energy as derived from the effective Hamiltonian. One therefore
expects them to be closer when interactions are less important. As we will
indeed see in Section~\ref{subsubsec:timesc_lowener}, at smaller coupling and
energy the mismatch becomes much smaller.

Looking at the contributions $E_{\text{mf}}$ and $E_{\text{modes}}$
to $H_{\text{eff}}$, we see a relatively rapid transfer
of energy from the mean field to the modes until a time of the order
$tm\approx50$.
This exchange takes place fairly locally in momentum space, as is found by
examining the mean field and mode contributions to $n_k$, a phenomenon that
we call local $k$-space equilibration.
At time
$tm\approx100$ most of the particle number already comes from the modes,
whereas the total distribution is still reasonably close to its initial form.
After $tm\approx50$,
energy is still going to the modes, but with a slower rate.
The behaviour in this second region,
from $tm\approx 50$ until $tm\approx 2000$, (see
Fig.~\ref{fig:ener_flat_medium}) can be fitted reasonably well with an
exponential form
\begin{equation}
A + B e^{-t/\tau},
\label{eq:expofunc}
\end{equation}
yielding $\tau m\approx100-150$.
\label{page:enerexchhigh} If we look at the long time behaviour, as plotted in
Fig.~\ref{fig:ener_flat_long}, we see there is also a much longer time scale
of the order $6000$, on which energy is going back into the mean field.
This time scale is comparable to the time scale of the emerging power-law
behaviour, discussed in Section~\ref{subsubsec:dist_highener}. The appearance of
this power law
is accompanied by
a large increase in the particle number in the zero mode
of the mean field and therefore also in the average energy density of the mean
field.
We recall that classical behaviour only becomes visible at
larger time scales of the order 15000.

In order to make a quantitative comparison between different couplings and
energies for the initial rapid exchange of energy between mean field and modes,
related to the local thermalization, we fitted the energy density in the
mean field to a function of the form \eqref{eq:expofunc}. The results are
summarized in Table~\ref{tab:flat_timesc}. Using the energy in the
quasi-particle picture, instead of the effective Hamiltonian, gives the same
results.
\begin{table}[tbp]
\begin{tabular}{c|c|c|c|}
$\tau m$ & $E/Lm^2=1$ & $E/Lm^2=2$ & $E/Lm^2=4$ \\
\hline
$\lambda/m^2=1/6$ & $137$ & $70$ & $39$ \\ 
$(tm<500)$ &&& \\
\hline
$\lambda/m^2=1/8$ & $215$ & $108$ & $50$ \\ 
$(tm<800)$ &&& \\
\hline
$\lambda/m^2=1/12$ & $688$ & $207$ & $112$ \\ 
$(tm<2500)$ &&& \\
\hline
\end{tabular}
\caption{\label{tab:flat_timesc} Initial energy-exchange time scales for the
flat-ensemble initial conditions.}
\end{table}

Leaving out the run at the lowest coupling and energy,
$\lambda/m^2=1/12, E/Lm^2=1$,
which we did not see thermalize, the time scale is roughly proportional to
$E^{-1}$ at constant coupling and to $\lambda^{-3/2}$
at constant energy
density:
\begin{equation}
\tau^{-1} \approx C m (E/Lm^2) (\lambda/m^2)^{3/2}.
\label{eq:timescale}
\end{equation}
We have checked this behaviour explicitly by plotting the different energies as
a function of
$(\lambda/m^2)^{3/2} (E/Lm^2)\, t$, which led to $C = 0.10$.

\begin{table}[tbp]
\begin{tabular}{p{0.24\textwidth}|c|c|}
$\tau m$ & Peak 1 \& 2 & Peak 1 \& 3 \\
\hline
\symmetric{},
Hartree & $160 \pm 31$ & $360 \pm 35$ \\
\hline
\symmetric{},
classical & $90 \pm 18$ & $156 \pm 34$ \\
\hline
\broken{},
Hartree & $49 \pm 11$ & $84 \pm 14$ \\
\hline
\broken{},
classical & $41 \pm 14$ & $63 \pm 16$ \\
\hline
\end{tabular}
\caption{\label{tab:flat_autocorr} Auto-correlation times for flat ensemble
type initial conditions. In all cases the coupling
$\lm/|\mu_{\text{ren}}^2| = 1/6$.
In the \symmetricph{} $v^2=0$, $\lm/m^2 = 1/6$ and
$E/Lm^2=4$, whereas in the \brokenph{} $v^2=6$, $\lm/m^2 = 1/12$,
and $E/Lm^2=0.5$.}
\end{table}

For the lower energy density $E/Lm^2= 0.4$ the results are very similar to our
simulation of the Gaussian wave packet, which we describe in
Section~\ref{subsec:lowener}.
In particular, we encountered the local $k$-space equilibration. 
If we initially excite only a few modes this process can be seen even more
clearly. We have not simulated long enough at this low energy density
using the flat initial distribution to see the emergence of classical behaviour.

\subsubsection{Auto-correlation time scales}

\begin{figure}[tbp]
\includegraphics[width=0.48\textwidth]{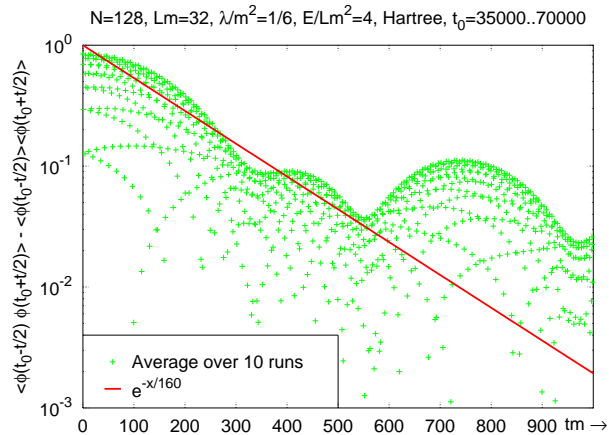}
\caption{\label{fig:autocorflat}
Auto-correlation function for the average mean field using the
flat Hartree ensemble.
The lower ``arcs'' are an effect of sampling the oscillations
$\propto \log |\cos m_{\text{eff}}t|$ at discrete times.}
\end{figure}

To further investigate time scales, we also analysed the
time-dependent auto-correlation function of the mean field,
as in \cite{SaSm00c,SaSm00d}. Using flat ensemble
initial conditions in both the \symmetric{} and the \brokenph{}, with either
Hartree or classical dynamics the auto-correlation function was obtained from
the average mean field only:
\begin{equation}
C(t) =
\overline{\langle\bar{\phi}(t_0-t/2)\bar{\phi}(t_0+t/2)\rangle}^{t_0} -
\textrm{d.c.}
\label{eq:autocorr}
\end{equation}
Here $\bar{\phi}(t)$ denotes the spatially averaged mean field, the long
overline 
includes averaging over a large time interval (which greatly reduces 
fluctuations), and d.c.~stands
for the disconnected piece.
Fig.~\ref{fig:autocorflat} shows an example in the large-time region,
where the particle-number distribution has the behaviour shown in
Fig.~\ref{fig:powerlaw}. The time average was taken over the region
$tm = 35000 \ldots 70000$,
and ten initial configurations from
the flat ensemble.
We recall that the evident damping is seen also upon
using only a single configuration \cite{SaSm00a}, it is not caused by
the average over initial conditions. The dip-like structure
can be understood as being caused by interfering ``twin peaks''
in the spectral function \cite{SaSm00c}.
The damping time is quantified using a ``fit'' of the form $\exp(-t/\tau)$
through
the first and second peaks. Using the third peak would give a roughly twice as
large $\tau$. The results are given in Table~\ref{tab:flat_autocorr}, where the
errors are obtained with the jackknife method~\cite{Mi74}. In the table a
comparison is made with results from the classical approximation, and with
results obtained in the \brokenph{}.

In the \brokenph{} there is hardly any difference between the classical and the
Hartree results, although the Hartree result seems to indicate a slightly larger
value. We recall that the particle distribution in this case approximates
the Bose-Einstein form reasonably well, and furthermore,
that the damping time is within
a factor of two of the analytically computed value using perturbative
quantum field theory in the two-loop approximation \cite{SaSm00a}.

In the \symmetricph{} the Hartree result is roughly twice the classical
value. It is hard to interpret this in any detail as the distribution function
in the Hartree case is so ``unconventional'' (cf.~Fig.~\ref{fig:powerlaw})
and also the classical case is far from thermalized.
However, there is a much more striking difference between the \broken{} and
\symmetricph{} results.
At an $8$ times larger energy,
the auto-correlation time in the \symmetricph{} is not smaller,
but instead larger by a factor of $3-4$.
One would expect qualitatively the opposite effect.
For example, for a thermalized system at a
temperature $T$ the damping rate may be expected to scale,
in the classical approximation, as $(\lm T)^{1/3}$, and bluntly using the
values $\lm T/m^3 = (1/6) (1/0.41)$ (\symmetric{}, Table \ref{tab:flat_temp})
and (1/12) 1.1 (\broken{}, \cite{SaSm00a}) would give
$\tau_{\text{\symm}}/\tau_{\text{\broken{}}} = 0.62$ instead of the factor
$3-4$.

Comparing with the time scale for energy exchange, we see that at high energy 
density the damping time is $4-9$ times larger than the energy-exchange time
(cf.~Table~\ref{tab:flat_autocorr} and the upper-right entry 
in Table~\ref{tab:flat_timesc}).
The systematics of this is unclear to us:
at the lowest energy density (and smaller coupling)
we find on the contrary that the damping time is about 
half the time scale for energy exchange 
(see also Sec.~\ref{autoGauss} for the Gaussian wave packet: 
$\tau_{\text{damp}} \approx 3500$, $\tau_{\text{exch}} \approx 7000$). 

\subsection{Gaussian wave packet \label{subsec:lowener}}

In this section we focus on the initial condition specified by
the Gaussian wave packet \eqref{eq:inigauss} with $\lambda/m^2 = 0.1$,
$Am^2=2$ and $\Phi=2.60106$ (this value appeared in the preprint version
of \cite{BePa01}), which gives an energy $E/m^2 = 12.6$.
We used a volume $Lm=32$, giving
an energy density $E/Lm^2 = 0.394$ which is practically equal to the smallest
energy density $0.4$ studied in the previous section
with the flat ensemble. It is however still
an order of magnitude
larger than the highest energy densities studied in \cite{BePa01}.
Our lattice size in this case was $N=256$ lattice points, and
the temporal lattice distance $a_0 = a/10$, as before.
We checked for finite volume and discretization effects
by using different parameters and found
that
they do not influence the results discussed.

\subsubsection{Particle distribution function \label{subsubsec:dist_lowener}}

The initial Gaussian wave packet spreads and oscillates in the course of
time and after $t\gtrsim L/2$ the packet meets itself
through the periodic volume.
This can be seen from the plots of the mean field $\ph(x)$,
see Fig.~1 in \cite{BePa01}, which we have verified.

The initial wave packet \eqref{eq:inigauss} represents a pure state, which
can still be analyzed in terms of particle numbers and frequencies obtained
from the two-point functions, as in Eqs.~\eqref{eq:quasipart}. 
It is interesting to compare the so-obtained $n_k$
with the coarse-grained particle distribution at later times.
If we assume free-field evolution we can calculate $n_k$
analytically, and it turns out that its average over (half) an oscillation
period is time independent and close to the initial distribution,
for large volumes.
As derived in Appendix~\ref{app:deriv2pnt}
this free-particle distribution for the Gaussian wave packet as
initial condition is given by
\begin{equation}
n_k^{\text{free}} = \frac{\pi A\Phi^2\sqrt{m^2 + k^2}\, e^{-k^2 A}}{L}.
\label{eq:dist_analyt}
\end{equation}

\begin{figure}[tbp]
\includegraphics[width=0.48\textwidth]{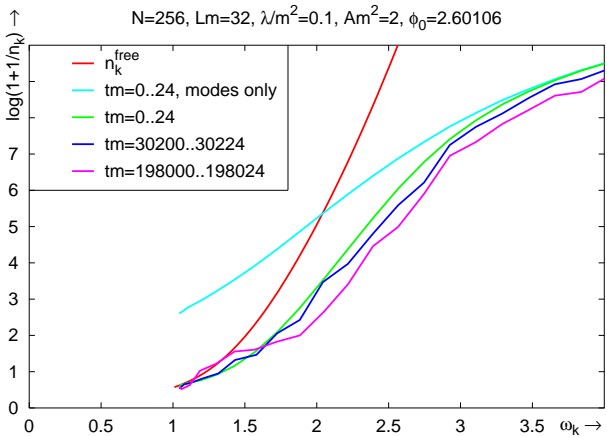}
\caption{\label{fig:dist_gauss_late}
Time development of particle number. The order in the key corresponds to
the curves from top to bottom at $\om_k/m = 2.5$.}
\end{figure}

In Fig.~\ref{fig:dist_gauss_late} we plot this free form together with
the particle numbers obtained in a simulation.

We find it quite remarkable that
already the earliest (time-averaged) distribution deviates significantly from
the initial form \eqref{eq:dist_analyt}.
A closer look shows that this deviation originates entirely from
the first period ($tm=0\dots2\pi$). After that short time the
distribution is almost stationary. Only after a time $tm=\mathcal{O}(10^5)$ do 
we see deviations arise. However,
in the mean time there is an extensive exchange of
energy between the modes and the mean field: initially all particle number and
energy is contained in the mean field, while in the later stage it is just the
opposite.

After $tm\approx30000-40000$ classical behaviour, i.e.~$n_k \to T/\omega_k$,
starts to emerge: the lower-momentum modes become under-occupied, while the
higher modes become over-occupied. At no stage does the distribution resemble
the Bose-Einstein form \eqref{eq:be}. We recall that also with the flat
ensemble we did not see quantum thermalization at similarly low energy
densities.

Bettencourt et al.~\cite{BePa01} studied the power spectrum of
the subtracted two-point function $S_k(t) - S_k(0)$ at times $tm \lesssim 200$.
This appeared to show power behaviour $\sim k^{-3}$ to $k^{-4}$,
which was interpreted as evidence for the absence of BE-like
thermalization. As mentioned above we also see no BE thermalization
at this low energy density, but
we find that the power behaviour is not without ambiguities.
The aim of the subtraction in $S_k(t) - S_k(0)$ was to eliminate the
vacuum contribution $1/(2\sqrt{m^2 + k^2})$ from $S_k$.
At large $k$ this is a rather delicate procedure. For instance,
a quasi-particle behaviour
$S_k(t) = (n_k(t) + 1/2)/\sqrt{m(t)^2 + k^2}$ with a thermal-like mass
$m(t)$ that is expected to be larger than
$m$ in the \symmetricph{}, would give a
negative result at large $k$,
$S_k(t) - S_k(0) \approx  - [m(t)^2 - m^2]/4k^3 $, where we neglected an
assumed exponentially small $n_k(t)$.
\begin{figure}[tbp]
\includegraphics[width=0.48\textwidth]{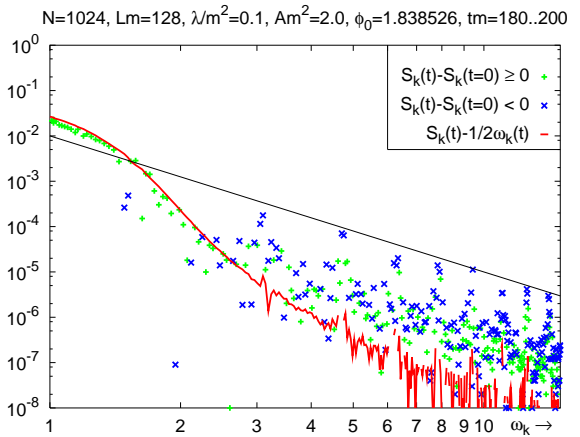}
\caption{\label{fig:power} The power spectrum $S_k(t) - S_k(0)$
and $n_k(t)-1/2\om_k(t) = S_k(t) - 1/2\om_k$ for $tm = 180 \cdots
200$. The line represents a power $\propto \om^{-3}$ touching the
negative values of $S_k(t)-S_k(0)$.}
\end{figure}

We would like to stress here the good features of the observables
$n_k$ and $\om_k$ defined in Eqs.~\eqref{eq:quasipart}. In Fig.~\ref{fig:power}
we have plotted $S_k(t) - 1/2\om_k(t) =
n_k(t)/2\om_k(t)$, as well as $S_k(t) - S_k(0)$, for the same
parameters ($\Phi = 1.838526$, $Am^2 = 2$, $Lm=128$, $N=1024$) and
in the same time regime as used in \cite{BePa01}. (We averaged
over $tm = 180-200$, which hardly affects $n_k(t)/2\om_k(t)$ as it
is practically constant.) The plot of $S_k(t) - S_k(0)$ looks very
similar to the ones shown in \cite{BePa01}. There is a lot of
scatter at large $\om_k$ ($\approx k$) and a more detailed
analysis shows negative values interspersed with positive values
(indicated separately for the power spectrum). On the other hand
$n_k/2\om_k$ shows less scatter and is mostly positive (only for
$\om_k > 4$ do negative values occur). However, note that the
larger $\om$ region could be affected by lattice artefacts.

\subsubsection{Energy densities and time scales
\label{subsubsec:timesc_lowener}}

\begin{figure}[tbp]
\includegraphics[width=0.48\textwidth]{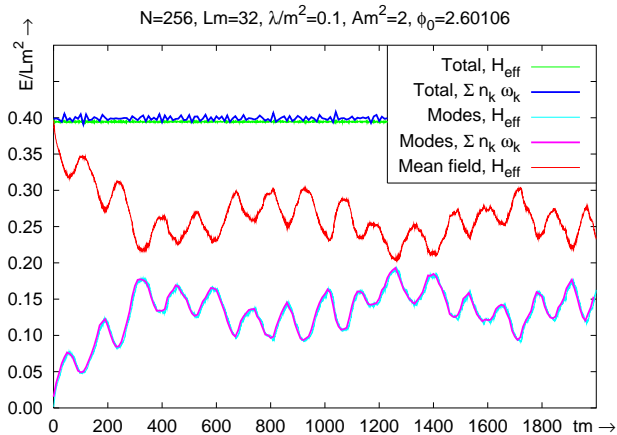}
\caption{\label{fig:ener_gauss_short} The different energies for the Gaussian
wave packet initial conditions at short times. From top to bottom: energy from
modes and mean field, energy from mean field, energy from modes.}
\end{figure}

\begin{figure}[tbp]
\includegraphics[width=0.48\textwidth]{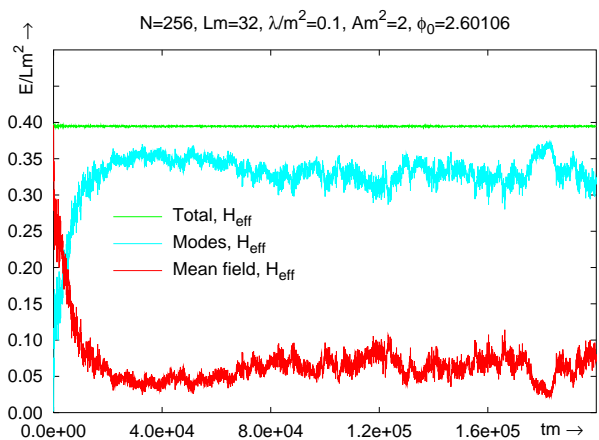}
\caption{The different energies for the Gaussian wave packet initial conditions
at long times.\label{fig:ener_gauss_long}}
\end{figure}

To get an estimate of the time scales involved we compare the energy
densities in the mean field, in the modes and in the total field 
for the Gaussian wave packet initial condition, as we did in 
Sec.~\ref{subsubsec:timesc_highener} for 
the flat ensemble at higher energy densities. For short times 
these are plotted in Fig.~\ref{fig:ener_gauss_short}, together with the
energy as derived from the quasi-particle picture \eqref{eq:partener}. For
long times they are plotted in Fig.~\ref{fig:ener_gauss_long}.

We see that the quasi-particle representation of the energies is in this case
extremely good, there is hardly any visible difference with the exact energies
based on $H_{\text{eff}}$.

Furthermore, for early times (Fig.~\ref{fig:ener_gauss_short}) there is an
oscillatory behaviour with a period $tm\approx 130$. Note that, due to the
periodic boundary conditions on the system with size $Lm=32$, the Gaussian
packet already ``meets itself'' after a time $tm=16$, much shorter than this
resonance time.
The resonance is caused by the difference between the effective mass terms
of the modes and mean field, $\approx 2\lambda \phi^2$.
This mass difference has a small value, fluctuating around 
$0.030-0.050$, 
corresponding to a period 
$210-126$, 
approximately the observed period.

The rate at which energy flows to the modes can be seen at long times
(Fig.~\ref{fig:ener_gauss_long}). The energy in the mean field in the interval
$tm<60000$
can be fitted reasonably well
to an exponential function of the form \eqref{eq:expofunc},
yielding an equilibration time scale
$\tau m \approx 7000$,
roughly two orders of magnitude larger than what was found in the \brokenph{}
at similar energy densities and couplings.

Using a sum of waves as initial condition at similar energy density
shows the same kind of resonance in the energy exchange between mean field and
modes, with period $\approx 170$ and mass difference
fluctuating around $0.02-0.04$.
Averaging over the flat ensemble the oscillations die out after
$tm\approx 4000-5000$.

\subsubsection{Time scales from the auto-correlation function\label{autoGauss}}

\begin{figure}[tbp]
\includegraphics[width=0.48\textwidth]{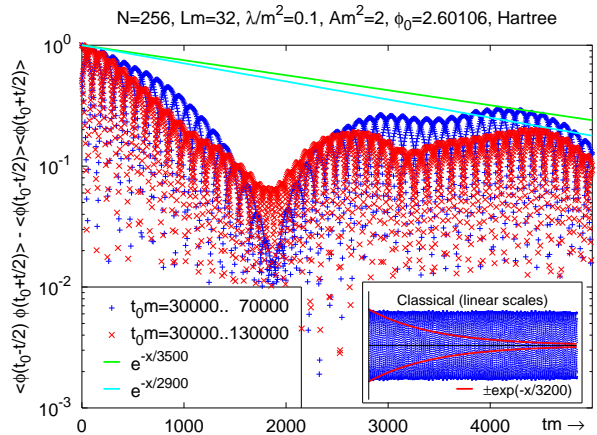}
\caption{The auto-correlation function as determined
from the average mean field only. The result when using
classical dynamics is shown in the insert on a linear scale, with
an exponential fit to the Hartree result. \label{fig:autocorr_gauss}}
\end{figure}
We also evaluated the auto-correlation function for the Gaussian wave packet.
Since we do not average over initial conditions we cannot calculate a
statistical error. We therefore averaged over two different time intervals,
giving some idea of the size of the statistical uncertainty. The result is
plotted in Fig.~\ref{fig:autocorr_gauss}. At this low energy, the damping time
is roughly half the energy equilibration time. The inset shows -- on a linear
scale -- the result for classical dynamics, using identical initial conditions.
The exponential curve is a fit to the Hartree result: classically there is no
visible damping.

It would be interesting to compare the result with the flat initial ensemble
at the same energy density. However, for these low energies we need
to simulate for very long times, which is quite a numerical effort.
We therefore only calculated the auto-correlation times for the
faster evolving high energy runs discussed in Section~\ref{subsec:highener}.

\section{Scattering\label{sec:scatter}}

In this section we will discuss
scattering features of the Hartree approximation.
First we take a fresh look at the possibility of scattering via
inhomogeneous mean fields by considering an initial state of
two localized particle wave packets in position space. 
Next we give a perturbative explanation of ``local $k$-space equilibration'',
our numerical result that 
the modes appear to equilibrate with the mean field primarily 
when they have the same wave number.  This occurs
especially at low energy density and weak coupling and it
has the effect that
the initial particle distribution in the mean field is 
taken over approximately by the modes.
We end this section with a brief discussion of higher-order 
scattering and thermalization.

\subsection{Scattering of two wave packets}

For homogeneous mean fields the Hartree approximation cannot describe true
scattering in which the momenta of the particles in the initial and final states
differ. However, in the inhomogeneous case the particles can truly scatter
off the mean field. It is interesting to take an intuitive look at this in
position space. Consider an initial two-particle state described by wave packets
$\ps_{1,2}$:
\begin{equation}
|\ps_1\ps_2\rangle = \bhd[\ps_1]\bhd[\ps_2]\vac, \quad
\bhd[\ps] = \sum_{\veck} \ps_{\veck} \bhd_{\veck}.
\end{equation}
This is a non-Gaussian and pure state, for which
\begin{equation}
C_{\text{ren}}(\vecx,t;\vecx,t) =  |\ps_1(\vecx,t)|^2 + |\ps_2(\vecx,t)|^2
\end{equation}
where
\begin{equation}
\ps(\vecx,t) = \sum_{\veck} \ps^*_{\veck} f_{\veck}(\vecx,t).
\end{equation}
Linearizing the Hartree equations in the \brokenph{},
writing $\ph = v + \ph'$ and keeping terms linear in $\ph'$ while
treating $|\ps|^2$ as being of the same order as $\ph'$, gives
\begin{eqnarray}
(\partial_t^2 - \Delta + m^2 ) \ph' &=& -3\lm \left(|\ps_1|^2 +
|\ps_2|^2\right), \nonumber\\
(\partial_t^2 - \Delta + m^2 + 6\lm v \ph') \ps_{1,2} &=& 0.
\end{eqnarray}
If the wave packets approach each other within a distance of order $1/m$ they
will scatter. 

So the interaction
of the quantum modes with the classical modes of the inhomogeneous mean field
does lead to indirect scattering.
Note that this happens especially in the \brokenph{}:
in the \symmetricph{} $v=0$ and the backreaction of
the mean field disturbance $\ph'$ to the particle waves $\ps_{1,2}$
is suppressed.

\subsection{Local $k$-space equilibration \label{subsec:localtherm}}

To give an analytic interpretation of the ``local $k$-space equilibration''
it is useful to focus on various interaction terms in the effective
Hamiltonian~\eqref{eq:effhamil}. Although derived from a quantum system, this
Hamiltonian can also be seen as describing interacting classical fields $\phi$
and $f_\alpha$. It will be convenient to split the modes $f_\alpha$ in a free
part and a perturbation:
\begin{subequations}
\begin{gather}
f_\alpha(x,t) = f_\alpha^0(x,t) + g_\alpha(x,t), \\
f_\alpha^0(x,t) = \frac{e^{ik_\alpha x -i \omega_\alpha t}}
{\sqrt{2 \omega_\alpha
L}},
\end{gather}
\end{subequations}
with $\omega_\alpha^2 = m^2 + k_\alpha^2$.
$f_\alpha^0$ will then play the role of an external field, as it is not
altered by the interaction.

We will show in the following that for not too large coupling and energy
the equation of motion for $g_\alpha$ reduces to that of a driven harmonic 
oscillator. Making use of the corresponding
scattering diagrams we then conclude that, approximately, 
the only momentum modes of $g_\alpha$
that are excited are those also present in the mean field.
Since we will focus on the
initial behaviour, when the system is still far from equilibrium and there is
not yet a temperature, we will only use zero-temperature perturbation theory.

We can write out the effective Hamiltonian in terms of the classical fields
$\phi$, $g_\alpha$ and the ``external field'' $f_\alpha^0$. In the
\symmetricph{} and to second order in $g_\alpha$, this leads to the following
interaction terms and corresponding vertex factors
\begin{subequations}
\begin{flalign}
\qquad  & \frac{1}{4} \lambda \phi^4
    & 6 \lambda & \qquad \label{eq:coupl_phi4} \\
&     3 \lambda \phi^2 \sum_\alpha \re(f_\alpha^0 g_\alpha^*)
    & 3 \lambda & \label{eq:coupl_phi2fg} \\
&     \frac{3}{2} \lambda \phi^2 \sum_\alpha |g_\alpha|^2
    & 3 \lambda & \label{eq:coupl_phi2g2} \\
&     6 \lambda \sum_{\alpha,\beta} \re(f_\alpha^0 g_\alpha^*)
        \re(f_\beta^0 g_\beta^*)
    & 3 \lambda & \label{eq:coupl_f2g2}
\end{flalign} \label{eq:coupl_4pnt}
\end{subequations}
whereas in the \brokenph{},
writing $\phi = v + \phi'$, we also have the three-point interactions
\begin{subequations}
\begin{flalign}
\qquad  & \lambda v \phi^{'3}
    & 6 \lambda v & \qquad \label{eq:coupl_phi3} \\
&     6 \lambda v \phi' \sum_\alpha \re(f_\alpha^0 g_\alpha^*)
    & 3 \lambda v & \label{eq:coupl_phifg} \\
&     3 \lambda v \phi' \sum_\alpha |g_\alpha|^2
    & 3 \lambda v & \label{eq:coupl_phig2}
\end{flalign} \label{eq:coupl_3pnt}
\end{subequations}

\begin{fmffile}{symmetric_feynfig}

In a first approximation we neglect the back reaction on the mean field and
assume it is just oscillating around its minimum as a superposition of waves:
\begin{equation} \label{eq:freephi}
\phi(x,0) = \sum_{i=1}^{i_{\rm max}} A_i \sin(\omega_{K_i} t)
\cos(K_i x - \psi_i)
\end{equation}
where $\psi_i$ are random phases and $\omega_{K_i} = \sqrt{m^2 + K_i^2}$.

The exact Hartree dynamical equation for the mode perturbation $g_\alpha(x)$ in
terms of its Fourier transform $g_{\alpha k}$ is given by
\begin{multline}
\label{eq:eomperturb}
(\partial_t^2 + \omega_k^2) g_{\alpha k} = -  3 \lambda
\int dx \left( \phi(x)^2 + C_{\text{ren}}(x) \right) \\
\times \left(\frac{e^{i (k_\alpha-k) x -i \omega_\alpha t}}
{\sqrt{2 \omega_\alpha}} +
      \frac{1}{L}\sum_{k'} e^{i (k'-k) x} g_{\alpha k'} \right).
\end{multline}
Neglecting for the moment the higher order terms containing $C_{\text{ren}}$ and
$g_{\alpha k}$ in the integral, the $x$ integration can be performed,
resulting in a sum over
plane waves. The equation is that of a driven harmonic oscillator
\begin{equation} \label{eq:drivharm}
(\partial_t^2 + \omega_k^2) g_{\alpha k}(t)
= \sum_j B_j e^{-i \Omega_j t},
\end{equation}
which leads to resonances that grow linearly in time for
$\omega_k^2=\Omega_j^2$. By inserting the explicit form \eqref{eq:freephi} into
Eq.~\eqref{eq:eomperturb} we find for each pair $K_i$, $K_j$ four different
resonance relations:
\begin{equation} \label{eq:omegreson}
\omega_k = \pm \omega_{\alpha} \pm \omega_{K_i} \pm \omega_{K_j},
\end{equation}
(with uncorrelated $\pm$),
while the $x$ integration gives four different momentum relations:
\begin{equation}
k=k_\alpha + \eta_1 K_i + \eta_2 K_j, \label{eq:momentrestric}
\end{equation}
where $\eta_{1,2}=\pm1$.

\begin{figure}[tbp]
\begin{center}
\fmfframe(0,5)(0,5){
 \begin{fmfgraph*}(50,50)
 \fmfleftn{i}{2}
 \fmfrightn{o}{2}
 \fmf{dots}{i1,v}    \fmfv{label=$k_\alpha$}{i1}
 \fmf{plain}{i2,v}   \fmfv{label=$K_i$}{i2}
 \fmf{dashes}{v,o1}  \fmfv{label=$k$}{o1}
 \fmf{plain}{v,o2}   \fmfv{label=$K_j$}{o2}
 \fmfv{label=$6\lambda$}{v}
 \end{fmfgraph*}
}
\qquad
\fmfframe(0,5)(0,5){
 \begin{fmfgraph*}(50,50)
 \fmfleftn{i}{2}
 \fmfrightn{o}{2}
 \fmf{dots}{i1,v}    \fmfv{label=$k_\alpha$}{i1}
 \fmf{plain}{i2,v}   \fmfv{label=$K_j$}{i2}
 \fmf{dashes}{v,o1}  \fmfv{label=$k$}{o1}
 \fmf{plain}{v,o2}   \fmfv{label=$K_i$}{o2}
 \fmfv{label=$6\lambda$}{v}
 \end{fmfgraph*}
}
\qquad
\fmfframe(0,5)(0,5){
 \begin{fmfgraph*}(50,50)
 \fmfleftn{i}{2}
 \fmfrightn{o}{2}
 \fmf{plain}{i1,v}   \fmfv{label=$K_j$}{i1}
 \fmf{plain}{i2,v}   \fmfv{label=$K_i$}{i2}
 \fmf{dashes}{v,o1}  \fmfv{label=$k$}{o1}
 \fmf{dots}{v,o2}    \fmfv{label=$k_\alpha$}{o2}
 \fmfv{label=$6\lambda$}{v}
 \end{fmfgraph*}
}
\end{center}
\caption{Tree level scattering diagrams
involving a single perturbation mode $g_{\alpha}$. 
Solid lines denote $\phi$, a dotted line $f_\alpha^0$, and the dashed line
denotes $g_\alpha$. Time runs from left to right.
\label{fig:localtherm}}
\end{figure}
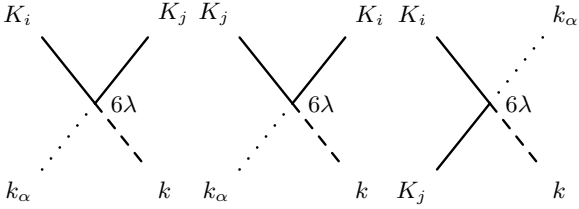
These two relations describe energy-momentum conservation in scattering
processes involving a single 4-point vertex, the
interaction~\eqref{eq:coupl_phi2fg}. Only $2\to2$ processes involving this
vertex can conserve energy and momentum. Furthermore, in $1+1$ dimensions (since
all particles have the same mass) it follows that the pair of incoming momenta
must be equal to the pair of outgoing momenta. From the energy
relation~\eqref{eq:omegreson} it then follows that there are three possible
diagrams,
drawn in Fig.~\ref{fig:localtherm}, creating a $g_\alpha$ particle with
momentum $k$. The momentum relation~\eqref{eq:momentrestric} now gives us three
possibilities,
\begin{subequations}
\begin{align}
k&=\eta_2 K_j &  k_\alpha &=-\eta_1 K_i \\
k&=\eta_1 K_i &  k_\alpha &=-\eta_2 K_j \\
k&= k_{\alpha}      &  \eta_1 K_i &= -\eta_2 K_j
\end{align}
\end{subequations}
For the last possibility $k=k_{\alpha}$, the $\phi(x)^2$ term contributes a
constant.
These terms only give rise to a time-dependent mass shift between the modes and
mean field. We can conclude that to leading order the only excited modes are
given by
\begin{equation} \label{eq:pert1order}
g_{\pm K_i,\pm K_j}.
\end{equation}
Note that only when just one mean field mode is excited (i.e. when
$i_{\text{max}}=1$) the modes will remain diagonal.

We will now look at the neglected terms. The renormalized mode sum
$C_{\text{ren}}(x)$ is equal to
\begin{equation}
C_{\text{ren}}(x) =
\sum_{\alpha}
\Bigl({f^0_\alpha}^*(x) g_\alpha(x) + f^0_\alpha(x) g_\alpha^*(x)
+|g_\alpha(x)|^2\Bigr).
\end{equation}
As we just showed, in lowest order, $g_\alpha$ is only non-zero for
$k_\alpha \in \{K_i\}$ and therefore the only non-zero Fourier components of
$C_{\text{ren}}(x)$ are the same as
those in $\phi(x)^2$: $k=\pm K_i \pm K_j$. Therefore including the first order
result for $C_{\text{ren}}(x)$ in Eq.~\eqref{eq:eomperturb}
will not change the set of excited modes.
Finally, including the last term in Eq.~\eqref{eq:eomperturb} using the first
order
result \eqref{eq:pert1order} we can also find its contribution.
The $x$ integration gives a $\delta_{k',k \pm K_i\pm K_j}$.
The frequencies of
the correction to
$g_\alpha$ are therefore of the form
$\omega_{k\pm K_i \pm K_j}$ and we find exactly
the same relation as following from Eqs.~\eqref{eq:omegreson} and
\eqref{eq:momentrestric}.

The above treatment can be extended by making a systematic expansion
in $\lambda$, $\phi = \phi_0 + \lambda \ph_1 + \lambda^2 \phi_2 + \cdots$,
$f_{\alpha} = f_{\alpha}^0 + \lambda f_{\alpha}^1 + \lambda^2 f_{\alpha}^2
+ \cdots$,
and using Green function techniques along the lines of \cite{AaSm97}.

As a check we performed a simulation, exciting only two modes $K_1$ and $K_2$ at
low energy ($E/Lm^2=0.04$) and small coupling ($\lambda/m^2=1/12$). The
assumption of a free oscillating mean field turned out to be extremely good. We
also checked the explicit form of one of the modes by examining $|f_{K_1}|^2$.
We expect $f$ to contain the two Fourier modes $K_1$ and $K_2$, and therefore
$|f|^2$ to contain momenta 
$2 K_1$, $2 K_2$, $K_1 + K_2$ and $K_1 - K_2$,
which were indeed the only modes found. 
In similar simulations at higher energy we found that the back reaction to 
$\phi$ became more important, but the set of excited modes remained the same.

\subsection{Higher order scattering \label{subsec:thermaliz}}

\begin{figure}[tbp]
\fmfframe(10,5)(10,5){
 \begin{fmfgraph*}(50,50)
 \fmfstraight
 \fmfleftn{i}{2} \fmfrightn{o}{4}
 \fmf{plain}{i1,v1}  \fmfv{label=$k_1$}{i1}
 \fmf{plain}{i2,v1}  \fmfv{label=$k_2$}{i2}
 \fmf{plain,label=$k'$}{v1,v2}
 \fmf{plain}{v1,o4}  \fmfv{label=$k_3$}{o4}
 \fmf{plain}{v2,o3}  \fmfv{label=$k_4$}{o3}
 \fmf{dots}{v2,o2}   \fmfv{label=$k_\alpha$}{o2}
 \fmf{dashes}{v2,o1} \fmfv{label=$k$}{o1}
 \fmfforce{(w,h/4)}{o1}
 \fmfforce{(w,h/2)}{o2}
 \fmfforce{(w,3h/4)}{o3}
 \fmfforce{(w/4,h/2)}{v1}
 \fmfforce{(5w/8,h/2)}{v2}
 \end{fmfgraph*}
}
\qquad
\fmfframe(10,5)(10,5){
 \begin{fmfgraph*}(50,50)
 \fmfleftn{i}{2} \fmfrightn{o}{4}
 \fmf{plain}{i1,v1}  \fmfv{label=$k_1$}{i1}
 \fmf{plain}{i2,v2}  \fmfv{label=$k_2$}{i2}
 \fmf{plain,label=$k'$}{v1,v2}
 \fmf{dashes}{v1,o1} \fmfv{label=$k$}{o1}
 \fmf{dots}{v1,o2}   \fmfv{label=$k_\alpha$}{o2}
 \fmf{plain}{v2,o3}  \fmfv{label=$k_4$}{o3}
 \fmf{plain}{v2,o4}  \fmfv{label=$k_3$}{o4}
 \fmfforce{(w/2,h/6)}{v1}
 \fmfforce{(w/2,5h/6)}{v2}
 \end{fmfgraph*}
}
\caption{Leading $2\to 4$ scattering diagrams creating a 
$g_\alpha^{k}$ particle in the \symmetricph{}.
The intermediate line represents the retarded Green function.
\label{fig:2_4_scattering}}
\end{figure}
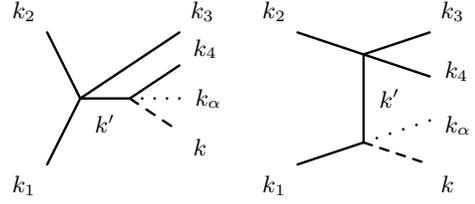
In order for the system to thermalize it is necessary that particles
can change their momenta by scattering. As mentioned above,
$2\to2$ scattering cannot change the initial momenta
in 1+1 dimensions.
(With re-summed off shell propagators this restriction does not apply, 
cf.~the nontrivial spectral functions in \cite{SaSm00c} and \cite{AaBe01},
and the thermalization found in \cite{BeCo01}.)
Therefore at least one extra vertex
is needed. In the \symmetricph{} only four-point vertices exist, as
in Eqs.~\eqref{eq:coupl_4pnt}.
The interaction~\eqref{eq:coupl_phi2fg} is leading over
\eqref{eq:coupl_phi2g2} and \eqref{eq:coupl_f2g2}, because it is
first order in $g_{\alpha}$, and since initially all energy is in the mean
field, the leading contribution to $g$-particle production comes from the two
diagrams in Fig.~\ref{fig:2_4_scattering}.

At this point it is interesting to realize what happens if the mean field is
homogeneous. In that case $g_\alpha$ always carries momentum $k_\alpha$. For
inhomogeneous systems this restriction is lifted and thermalization becomes
possible.

\begin{figure}[tbp]
\fmfframe(10,5)(10,5){
 \begin{fmfgraph*}(50,50)
 \fmfstraight
 \fmfleftn{i}{2} \fmfrightn{o}{3}
 \fmf{plain}{i1,v1}  \fmfv{label=$k_1$}{i1}
 \fmf{plain}{i2,v1}  \fmfv{label=$k_2$}{i2}
 \fmf{plain,label=$k'$}{v1,v2}
 \fmf{plain}{v2,o3}  \fmfv{label=$k_4$}{o3}
 \fmf{dots}{v2,o2}   \fmfv{label=$k_\alpha$}{o2}
 \fmf{dashes}{v2,o1} \fmfv{label=$k$}{o1}
 \fmfforce{(w/4,h/2)}{v1}
 \fmfforce{(3w/4,h/2)}{v2}
 \end{fmfgraph*}
}
\qquad
\fmfframe(10,5)(10,5){
 \begin{fmfgraph*}(50,50)
 \fmfleftn{i}{2} \fmfrightn{o}{3}
 \fmf{plain}{i1,v1}  \fmfv{label=$k_1$}{i1}
 \fmf{plain}{i2,v2}  \fmfv{label=$k_2$}{i2}
 \fmf{plain,label=$k'$,label.side=left}{v1,v2}
 \fmf{dashes}{v1,o1} \fmfv{label=$k$}{o1}
 \fmf{dots}{v1,o2}   \fmfv{label=$k_\alpha$}{o2}
 \fmf{plain}{v2,o3}  \fmfv{label=$k_4$}{o3}
 \fmfforce{(w/2,h/4)}{v1}
 \fmfforce{(w/2,3h/4)}{v2}
 \end{fmfgraph*}
}
\caption{Leading $2\to3$ scattering diagrams creating a $g_\alpha^{k}$ particle
in the \brokenph{}.\label{fig:2_3_scattering}}
\end{figure}
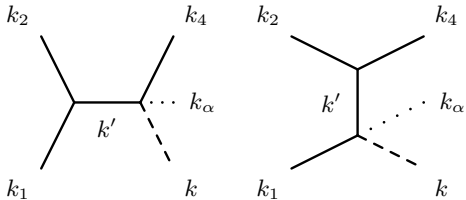
In the \brokenph{}
both the couplings~\eqref{eq:coupl_4pnt} and ~\eqref{eq:coupl_3pnt} contribute
and there are three- and four-point interactions.
The
leading contribution to $g$-particle production in this case comes from the two
diagrams in Fig.~\ref{fig:2_3_scattering}.
Intuitively one expects the finite range of the interaction in the \brokenph{},
due to off-shell particle exchange, to lead to more efficient
thermalization than the zero range interaction in the \symmetricph{}.
This is indeed what we observed.

\end{fmffile}

\section{Discussion \label{sec:discus}}

We will start this discussion with a summary of the behaviour at high
and low energy density. This appears to be the distinguishing
criterion for the thermalization behaviour of the Hartree approximation
for inhomogeneous systems, rather than, for example, an initial state being
pure, as for the Gaussian wave packet, or mixed, as for the flat ensemble.

At high energy density $E/Lm^2\gg 1$ we see that the distribution $n_k$ acquires
features of a thermal quantum, i.e.~Bose-Einstein distribution. There is a
time- and coupling-dependent chemical potential, of order unity in mass units.
The temperature is roughly proportional to $\sqrt{E/L}$ and independent of the
coupling. The coupling determines the time scale on which the approximate
thermalization becomes visible. The initial rapid exchange of energy between
modes and mean field occurs on a time scale described by
Eq.~\eqref{eq:timescale}.
The quasi-particle picture is reasonable
and the total particle number is very constant, in correspondence with the
chemical potential. However, there is a mismatch
between the total energy as derived from the effective Hamiltonian and that
obtained from the quasi-particles, Eqs.~\eqref{eq:effhamil} and
\eqref{eq:partener}. As we already pointed out, this mismatch depends on the
initial conditions, for example a Bose-Einstein type initial conditions
cf.\ Ref.~\cite{SaSm00b} leads to a much smaller mismatch. For an interacting
theory it is not expected that a quasi particle picture based on free particles
would give precisely the same value as what follows from effective Hamiltonian
and indeed the mismatch was practically invisible at lower energy and coupling.
Furthermore, a considerable amount of the total energy comes from the zero mode,
about $10\%$, making it very 
sensitive to the precise particle distribution
and probably making deviations from the quasi-particle picture more pronounced. 

After a very long time the energy flows back into the mean field, accompanied
by the emergence of a power-law distribution for $n_k$ as a function
of momentum $k$. Such power-law behaviour has not been found in the \brokenph{},
and also classical simulations do not show such a behaviour, indicating
that the back reaction of the modes plays an essential role. It is interesting
to note that Boyanovsky et al.~\cite{BoDe97} also found power-law behaviour for
the occupation numbers, although it is unclear if the same mechanism is behind
their finding. In their study of a $3+1$ dimensional $O(N)$ model in the large
$N$ approximation, the power-law is caused by a nonlinear resonance of the back
reaction of the modes on themselves, with terms of the form $1/(\omega_k-m)$,
diverging as $1/k^2$ in the limit $k \to 0$. This would give a $1/k^4$ behaviour
for the particle number, different from what is found here. Furthermore, we only
find power-law behaviour in the total- and mean-field-particle numbers, but not
in that of the modes. The difference in power and the absence of power-law
behaviour in the modes makes it improbable that the physical mechanism behind
the resonances is the same.

At low energy density $E/Lm^2 \ll 1$, for the flat ensemble as well as for the
pure-state wave packet, we do not find approximate thermalization to a BE
distribution. Instead, the form of the total distribution $n_k$ remains the same
for times that are many tens of thousands in units of $m^{-1}$. The distribution
then slowly turns over into a classical distribution. However, there is still an
extensive exchange of energy between the mean field and the modes, leading to
what we call local $k$-space equilibration, for which we were able to find an
interpretation based on a perturbative calculation.
At low energy, the time scale for this process is much longer than would follow
from Eq.~\eqref{eq:timescale}, found at high energy densities. Furthermore, at
short times the energy densities in the mean field and modes separately show a
remarkable oscillatory behaviour, not seen at higher energies, which is caused
by a difference in the effective mass of the mean field and modes.

We obtained several time scales:
for approximate Bose-Einstein thermalization,
for the early-time exchange of energy between mean field and modes, for the
auto-correlation function and for the evolution to a classical distribution.
Most of these are much longer in the \symmetricph{} than in the \brokenph{},
due to the absence of the finite range interaction.

According to \cite{SaSm00b}, the BE-thermalization-time scale in the \brokenph{}
is of the order $25-35$ for $E/Lm^2=0.5$,
while here, in the \symmetricph{}, $m\tau_{\text{BE}}=1500-1600$ for $E/Lm^2
= 4$ (both at $\lambda/|\mu_{\text{ren}}^2| = 1/6$).

The energy-exchange time scale in the \brokenph{} gives a result that is close
to $\tau_{\text{BE}}$,
whereas in the \symmetricph{} it is
related to local $k$-space equilibration,
much shorter than $\tau_{\text{BE}}$,
and it shows the behaviour \eqref{eq:timescale}.
For $E/Lm^2 = 0.5$ and
$\lambda/|\mu_{\text{ren}}^2| = 1/6$, Eq.~\eqref{eq:timescale}
gives $\tau m \approx 300$,
much longer than the $25-40$ we found in the \brokenph{} at the same energy
and coupling.

Also the damping time, obtained from auto-correlation functions, is much longer
than in the \brokenph{}, even at much higher energy and larger coupling.
Compared to the value obtained using classical dynamics, it is roughly twice as
large. In the \brokenph{}, both values are comparable in size. At low energies
the damping time seems to be much longer, but this needs more study.

The last time scale is that of classical equilibration. Since we are just
solving a large number ($2N^2 + 1$) of local classical non-linear equations, one
may expect classical equipartition to set in at some point. This equipartition
is, however, non-trivial because of the large number of conserved charges
\cite{SaSm00a}. For example, the emerging classical temperature is of order
$E/N$ and not $E/N^2$ \cite{SaSm00a}.\footnote{For $N$ spatial lattice sites
there are $2N^2$ real degrees of freedom in the mode functions.} Depending on
energy and coupling we can already see a first emergence of classicality at
times $\tau m=\mathcal{O}(10^4)$. This is still about an order of magnitude
longer than what was found in the \brokenph{} in \cite{SaSm00b}. However, full
classical equilibrium is expected only for huge times, much larger than the
$\tau m=\mathcal{O}(10^6)$ found in the \brokenph{} for an artificially small
system at $E/Lm^2 = 36$ \cite{SaSm00a}, and beyond the already large times of
order $10^5$ reached in this study.

Remarkably, the equilibration time scale found in \cite{AaBo00} using classical
dynamics appears to be {\em shorter}. The empirical formula \cite{AaBo00}
\begin{equation}
1/m \tau_{\class}=5.8~10^{-6}
(6 \lambda T/m^3)^{1.39},
\label{eq:clasdamp}
\end{equation}
with $T= E/N$ the classical equilibrium temperature, would give equilibration
times $tm=$ $\mathcal{O}(10^5)$ -- $\mathcal{O}(10^7)$ for the various
parameters used here. This difference in time scales can be interpreted as
follows.
Classical dynamics has also be studied in the Hartree approximation,
and the latter shows up as an unstable fixed point of the full
dynamics \cite{AaBo00}.
This Hartree fixed point depends on the initial conditions.
In our case the mode functions are
initialized with quantum-vacuum form \eqref{eq:inimode}, and the resulting
dynamics (seen as a classical system with order $N^2$ fields) appears to
linger for a very long time near a Hartree fixed point, longer than when
using classical dynamics.

For our inhomogeneous
initial conditions we have not been able to pin down the fixed point
analytically, but intuitively one may expect the system
to be close to it when the
mean field has lost most of its energy and has started fluctuating about
a homogeneous average. Making a homogeneous approximation to this situation
would lead to a Hartree stationary state.
Such a state can have an arbitrary particle distribution $n_k$, which,
given our out-of-equilibrium initial conditions,
turns out to have BE features when the energy density $E/Lm^2 \gg 1$.
Apparently, when the energy density is small, $E/Lm^2 \ll 1$, the system
leaves the fixed-point region before BE-like thermalization sets in, because
we have seen only classical-like equilibration emerging in this case.

Finally, we comment on the results of
Bettencourt et al.~\cite{BePa01}.
As mentioned in Sec.~\ref{subsubsec:dist_lowener},
we have essentially confirmed their numerical results.
The energy density in the simulations in \cite{BePa01} was rather low,
namely $E/Lm^2 = 0.00042$ and  0.0045,
so in view of our results summarized above, no sign of a BE distribution
is to be expected with the Hartree approximation at the times
$tm \lesssim 200$ covered in \cite{BePa01}, nor at any time later.

It is remarkable that we do find Bose-Einstein behaviour at larger energies
$E/Lm^2 \gg 1$, but of course, the fact remains that
one needs to improve on the Hartree approximation
in order to achieve thermalization at all energies. This may take huge times at
low energy densities.

It has been remarked \cite{BePa01} that the Hartree approximation is
expected to be valid up to times $tm \sim m^2/\lambda = {\cal O}(10)$.
We agree with this statement when applied to the detailed time-evolution
of observables, but it does not necessarily apply to observables such as our
quasi-particle distribution $n_k(t)$ or energy $\om_k(t)$,
which are coarse grained in time and space and/or averaged over initial
conditions in the Hartree ensemble approximation. For comparison, consider
a gas of classical point particles with Lennard-Jones interactions. Any
numerical approximation to the detailed time evolution will soon go dismally
wrong due to the chaotic nature of the system, but this does not preclude
an accurate evaluation of, say, a coarse-grained particle-distribution function.
With this in mind we have studied our system for times as large as seemed
necessary, which led to very large times indeed.
First experience \cite{JCV} indicates that the situation is not very different
in 3+1 dimensions, where also large equilibration times may be expected for the
$\phi^4$ model at moderate couplings and energy densities.

\acknowledgments{We thank Jeroen Vink for his collaboration in the
initial stages of this work. We thank Gert Aarts, Anders Tranberg and
Michele Simionato for useful discussions.
This research was supported by FOM/NWO. }

\appendix

\section{Particle number of the Gaussian wave packet
\label{app:deriv2pnt}}

We calculate here the initial two-point functions for the Gaussian wave packet
initial condition \eqref{eq:inigauss}, the corresponding particle number $n_k$
and energy $\om_k$, and their subsequent free field expressions. The
calculations will be made in the continuum limit, in our finite periodic volume.

The mean field contributions to the two-point functions are given by
\begin{subequations}
\begin{eqnarray}
S(x,y)^{\mf} &=& \overline{\phi(x) \phi(y)}
- \overline{\phi(x)}\;\;\overline{\phi(y)}, \\
U(x,y)^{\mf} &=& \overline{\pi(x) \pi(y)}
- \overline{\pi(x)}\;\;\overline{\pi(y)},
\end{eqnarray}
\end{subequations}
where at first we shall average only over space, i.e.
\begin{equation}
\overline{\phi(x) \phi(y)} = \frac{1}{L}\int_0^L dz\, \phi(x+z)\ph(y+z).
\end{equation}
The initial mean field is given by Eq.~\eqref{eq:inigauss}, or in terms of its
Fourier transform:
\begin{equation}
\phi_k  = \int dx\, e^{-i k x} \phi(x) =
\Phi\sqrt{2 \pi A}\; e^{-k^2 A/2}.
\label{eq:mf_fourier0}
\end{equation}
Since $\pi(0)=0$, the free-field
(i.e.~for $\lambda\to 0$ and $\mu\to \mu_{\text{ren}}= m$)
evolution of $\phi_k$ is given by:
\begin{equation}
\phi_k(t) = \phi_k(0) \cos(\omlat{0} t),
\label{eq:mf_fourier}
\end{equation}
where $\omlat{0}=\sqrt{m^2+k^2}$.
A straightforward calculation gives
\begin{subequations}
\begin{eqnarray}
S_k^{\mf} &=& \Bigl( 1 - \delta_{k,0} \Bigr)
\frac{\phi_k^2 \cos^2(\omlat{0} t)}{L}, \\
U_k^{\mf} &=& \Bigl( 1 -  \delta_{k,0} \Bigr)
\frac{(\omlat{0})^2\phi_k^2 \sin^2(\omlat{0} t)}{L},
\end{eqnarray}
\label{eq:mf2pnt}
\end{subequations}
where the delta functions come from the disconnected pieces.
The modes just contribute the vacuum fluctuations:
\begin{equation}
S_k^{\text{modes}} = \frac{1}{2 \omlat{0}}, \qquad
U_k^{\text{modes}} = \frac{\omlat{0}}{2}.
\label{eq:mode2pnt}
\end{equation}
Adding the contributions in Eqs.~\eqref{eq:mf2pnt} and \eqref{eq:mode2pnt}
and applying the definition \eqref{eq:quasipart},
the initial instantaneous particle number and frequency become
\begin{subequations}
\begin{eqnarray}
n_k(0) &=& \frac{1}{2} \left(\sqrt{2 \omlat{0} \phi_k^2 / L+1} - 1 \right),
\label{eq:partgauss}
\\
\omega_k(0) &=& \frac{\omlat{0}}{\sqrt{2 \omlat{0} \phi_k^2 / L +1}}.
\label{eq:omegauss}
\end{eqnarray}
\end{subequations}
Using free field dynamics the instantaneous particle number would
get an oscillating component according to Eqs.~\eqref{eq:mf2pnt}.
If we also course grain in time, the disconnected parts of $S$ and $U$
vanish, while both $\cos^2$ and $\sin^2 \to 1/2$. We then find
\begin{equation}
n_k^{\text{free}} = \frac{\omlat{0} \phi_k^2}{2 L}, \qquad
\omega_k^{\text{free}} = \omlat{0},
\label{eq:partomegfreegauss}
\end{equation}
which are time independent.

For large volumes
$2 \omlat{0} \phi_k^2 / L \ll 1$, expressions \eqref{eq:partgauss} and
\eqref{eq:omegauss} reduce to \eqref{eq:partomegfreegauss}. For the
parameters as used in Section~\ref{sec:numres}, $A=2$, $\Phi=2.60106$,
$Lm=32$, we have
\begin{equation}
\frac{2 \omlat{0} \phi_k^2}{L}
\approx 5.3 \sqrt{1 + k^2/m^2}\, e^{-2 k^2/m^2}.
\label{eq:nk_iniapprox}
\end{equation}
Plotting $n_k(0)$ (or $\log(1+1/n_k(0)$) versus $\om_k(0)$
we find that this only compares well with  a similar plot of
$n_k^{\text{free}}$ versus $\om_k^{\text{free}}$ for $k\gtrsim 2 m$.
So at times $tm \gg 1$
it is best to use the time-averaged free-field determinations
for the comparison with the interacting Hartree evolution.

\bibliography{liter}

\end{document}